\documentclass{aastex631}
\usepackage{amsmath,esint}

\newcommand*\dif{\mathop{}\!\mathrm{d}}
\usepackage{bm}

\shorttitle{AASTeX v6.31 Sample article}
\graphicspath{{./}{figures/}}
\usepackage{hyperref}
\DeclareRobustCommand{\rchi}{{\mathpalette\irchi\relax}}
\newcommand{\irchi}[2]{\raisebox{\depth}{$#1\chi$}} 

\begin{document}
\title{A numerical approach using a finite element model to constrain the possible interior layout of (16) Psyche}

\author[0000-0002-9042-408X]{Yaeji Kim}
\affiliation{Department of Aerospace Engineering\\ 
Auburn University, 211 Davis Hall\\
Auburn, Alabama 36849, USA}

\author[0000-0002-1821-5689]{Masatoshi Hirabayashi}
\affiliation{Department of Aerospace Engineering\\ 
Auburn University, 211 Davis Hall\\
Auburn, Alabama 36849, USA}
\affiliation{Department of Geosciences\\ 
Auburn University, 2050 Beard Eaves Coliseum\\
Auburn, Alabama 36849, USA}

\begin{abstract}
Asteroid (16) Psyche (278$\times$238$\times$171 km in size) is notable for the largest M-type asteroid and has the high radar albedo (0.34$\pm$0.08) among the main-belt asteroids. The object is likely a mixture of metal and silicates because of its lower bulk density ($\sim$4.0 g cm$^{-3}$) than metallic materials ($\sim$7.5 g cm$^{-3}$) and observations inferring the existence of silicate materials on the surface. Here, we numerically investigate the interior layout when the structure of Psyche consists of a spherical iron core and two types of the silicate-rich layers (compressed and uncompressed ones) resulting from the compaction process (later known as a Three-layer model). We develop an inverse problem algorithm to determine the layout distribution by combining a Finite Element Model (FEM) approach that accounts for density variations and constrains pressure-based crushing conditions. The results show that given the crushing limit of 10 MPa the smallest core size likely reaches 72 km in radius, and the silicate-rich layer, consisting of both compressed and uncompressed regions, has a thickness ranging up to 68 km. To support the localized metal concentration at the crater-like region detected in the recent radar observation, we give more constraints on the minimum core size which takes up to 34 - 40 $\%$ of the entire size of Psyche. Our study also addresses that the ferrovolcanic surface eruptions could still be a source of metal-rich materials. Finally, while the differentiated structure having a spherical core condition is just part of potential scenarios, the present study infers that the core and compressed layer conditions likely control the surface condition. Further investigations will provide key information for observable properties on NASA's Psyche mission to provide insight into its evolution history.
\end{abstract}


\section{Introduction}
\label{sec:intro}
Asteroid (16) Psyche is one of the largest M-type asteroids with a size of $\sim$250 km in the main asteroid belt. The object has long been discussed as a remnant of the iron core from an early planet \citep{zellner1985eight,chapman1979reflectance,gaffey1989reflectance,gaffey1993asteroid, binzel1995rotationally} and thus considered a unique relic originated from a differentiated planetesimal that can provide insights into the following questions about the formation of our Solar System: how an early planet has formed and evolved and where iron meteorites originated. To explore this object, NASA selected it as a target of its Discovery mission \citep{elkins2020observations,oh2016psyche} with a main scientific objective of determining whether Psyche is a core from a differentiated body \citep{yang2007iron,johnson2020ferrovolcanism}.

After Psyche was detected, the planetary science community dominantly thought that this body is a pure-metallic body \citep{ostro1985mainbelt,shepard2010radar,magri2007radar} because the observations measured significantly higher radar albedo ($\sim$0.37) than other main-belt asteroids, commonly S- and C- types (0.14 - 0.15) \citep{magri2007radar}. However,if the asteroid is a pure metal or dominantly metal-rich object, the most puzzling physical parameter is its bulk density. While the density of iron meteorites is $\sim$7.5 g cm$^{-3}$ \citep{smyth1995crystallographic}, the recent estimates of the Psyche's bulk density are all close to $\sim$4.0 g cm$^{-3}$ \citep{carry2012density,hanuvs2017volumes, drummond2018triaxial,viikinkoski201816,shepard2021asteroid,siltala2021mass,elkins2020observations}, which is remarkably lower than the expected value. This inconsistency may be understandable if Psyche is a rubble-pile asteroid having a very high porosity ($>$  $\sim$50$\%$). Also, \cite{nicholsporosity} recently indicated that Psyche might be difficult to reconcile with a fully metallic structure by showing that high porosity ($\sim$40$\%$) cannot be sustainable due to the porosity removal by thermal evolution. They found that a Psyche-sized body must cool down below 800 K prior to porosity-adding events (i.e., disruptive impacts) to persist the sufficient porosity ($\sim$40$\%$) until the present. However, the required timescale to cool down below 800 K (100s of Myr) is not compatible with the disruptive impacts (i.e., hit-and-run collisions \citep{asphaug2006hit}), which are most likely to have occurred within a few million years ($<$10 Myr) post-CAI formation \citep{yang2007iron}, although some studies of planet formation suggested that hit-and-runs might have occurred for 100s of Myr in the solar system (e.g., \cite{chambers2013late,emsenhuber2020realistic}). If the pure-metallic structure with high porosity is not the case for Psyche, we may interpret that this object has mixed silicates and metal. \cite{shepard2017radar, shepard2021asteroid} showed that radar albedo has significant variations across the entire surface as Psyche rotates. While the highest albedo follows that of M-type asteroids, the lowest value ($\sim$0.2) is within the high radar albedo ranges of S- and C-type asteroids. This variation indicates that Psyche does not have a constant composition across the surface. The recent works (e.g., \cite{shepard2021asteroid} and \cite{ferrais2020asteroid}) also support this conclusion. In addition, \cite{sanchez2016detection} and \cite{takir2016detection} showed the presence of orthopyroxene and hydrated silicates on the surface although they do not provide strong constraints on abundances of silicates. In advance, Psyche's thermal emission observations can support the existence of a silicate layer but still do not constrain its abundance. In terms of the thermal inertia, there were three findings: $100-150$ J m$^{-2}$ s$^{1/2}$ K$^{-1}$ by \cite{matter2013evidence}, $5-25$ J m$^{-2}$ s$^{1/2}$ K$^{-1}$ by \cite{landsman2018asteroid}, and $180-380$ J m$^{-2}$ s$^{1/2}$ K$^{-1}$ by \cite{de2021surface}. Since high thermal inertia generally indicate the abundance of metallic contents, \cite{matter2013evidence} and \cite{de2021surface} interpreted their measurements as the surface of Psyche contains abundant metal contents. Conversely, \cite{landsman2018asteroid} obtained a considerably lower value of thermal inertia that indicates the surface is highly likely to be covered with fine-grained silicates. The most recent work, \cite{de2021surface}, derived the spatially resolved thermal emission measurements further to discuss the possible compositions for Psyche's surface. This study concluded that the pure silicate and pure metal are unlikely for the Psyche's surface composition. Instead, the metal content would exist at least 20$\%$ on the surface.

Given that Psyche has metallic and silicate-rich components, a few scenarios are possible for its formation process. If Psyche originates from a differentiated planetesimal, it could be a fragmented remnant of the core with not an entirely stripped silicate layer \citep{davis1999missing,asphaug2014mercury}. \cite{asphaug2014mercury} performed numerical hydrocode simulations showing that proto-Mercury could remove its mantle layer via high-speed collisions with a larger target body that can support Psyche being a relic of inefficient accretion. With the assumption that Psyche is a differentiated silicate-iron body, the ferrovolcanic surface eruption proposed by \cite{abrahams2019ferrovolcanism} and \cite{johnson2020ferrovolcanism} is also applicable to explain potential high metal-rich components on the surface. Depending on whether the iron core includes sulfur-rich contents and the thickness of the silicate layer, Psyche might experience ferrovolcanism that erupted metallic components in the iron core onto the surface \citep{johnson2020ferrovolcanism}. The recent radar observation analysis by \cite{shepard2021asteroid} suggested that Psyche could be a differentiated object containing the surface layer with a regolith composed of similar enstatite or CH/CB chondrites and localized high metal concentration. They picked ferrovolcanism as the most credible formation mechanism, although an impact cratering process might cause the localized metal concentration at crater-like regions. Lastly, we address that it is still possible that Psyche might come from an unmelted/undifferentiated body. The meteorite having the best-fit density of Psyche is the CB Chondrites \citep{hardersen2011m} known as the most primitive meteorite group. However, if the primordial body is the case for Psyche, this interpretation raises questions about this body's origin \citep{elkins2020observations}. Furthermore, it is also questionable that its surface can induce the significant radar albedo variations detected in \cite{shepard2017radar,shepard2021asteroid}. Given that both scenarios for the differentiated and primordial structure are similarly possible, this study adopts the former as the working assumption and investigates the interior condition of the differentiated Psyche.

The data currently available for Psyche is still not enough to provide details on its interior. We thus refer to the structural condition of some small bodies that can provide insights. Most small-sized objects between $\sim$200 m to 10 km are dominantly rubble piles - gravitational aggregates - as seen in Itokawa ($\sim$330 m in diameter) \citep{fujiwara2006rubble}, Ryugu ($\sim$870 m in diameter) \citep{watanabe2019hayabusa2}, and Bennu ($\sim$480 m in diameter) \citep{dellagiustina2019properties}. If the object is much more tiny (less than a few hundreds or tens of meters), it is likely to be monolithic or strength-dominated body. Rather, the bigger sized object ($>$ tens of kilometers) is more likely to be shattered body whose structure is intact but fractured - as seen in Eros ($\sim$34 km in diameter), Gaspra ($\sim$12 km in diameter), and Ida ($\sim$32 km in diameter). For the larger asteroids ($> \sim$100 km), the one such as Vesta ($\sim$500 km in diameter) has a structure differentiated into the core, mantle, and crust \citep{keil2002geological}. Although it is unable to constrain where Psyche is subgrouped, we expect this asteroid might be in the transition between shattered and differentiated body only considering its size ($\sim$250 km in diameter). 

In this study, we model the structure of Psyche as layered by assuming that Psyche could originate from a differentiated planetesimal. The identical interior layout is a metallic core covered with a silicate-rich layer. With the assumption that the metallic core is spherical and the silicate-rich layer (later separated into two types: dense and less dense ones) overlays it, we numerically investigated the layer distribution (i.e., the size of iron core and thickness of the silicate-rich layer) by considering reasonable geophysical conditions on each layer. For this, we developed an inverse problem algorithm using a Finite Element Model (FEM) that provides stress fields of Psyche's structure by accounting for its density variations and pressure-based crushing limit in the silicate-rich layer. Our study eventually indicated that Psyche is still eligible to expose the metallic components in crater-like regions via an impact cratering process or experience ferrovolcanism in localized regions when the differentiated Psyche with the spherical core shape has a certain amount of silicates compatible with the reported bulk density of 4.0 g cm$^{-3}$.

\subsection{Compaction mechanism in the silicate-rich layer}
\label{subsec:compaction}
If the iron-core covered with the silicate-rich layer is the case for Psyche, the silicate-rich layer is likely to be compressed under high-pressure. When high pressure is applied, grains could be crushed, resulting in a more compact configuration. Earlier work (e.g., \cite{britt2003asteroid,hagerty1993one,yamamuro1996one}) described that, in the case of silicate grains, the grain fracture begins when the applied pressure exceeds $\sim$10 MPa. Especially for Psyche, the first few sub kilometers from the surface reach this crushing limit ($\sim$10 MPa). This allows the subsurface to hold lower porosity than the top surface layer. \cite{shi2016modeling} conducted one-dimensional compression tests using a discrete element method (DEM) model to study the particle breakage effect in silica sands. The simulations were set with the particle size ranging from 1.4 to 1.7 mm and a grain density of 2.6 g cm$^{-3}$, which is compatible density with the silicate layer of Psyche although we cannot constrain grain sizes yet. Their work demonstrated three stages in the compaction mechanism. The first stage is when the grains exceed the yield point ($\sim$10 MPa) but less than the maximum compression index ($\sim$20 MPa). In this stage, the particles start to be rearranged and the limited number of particles fractured, allowing the initial porosity to decrease up to $\sim$30$\%$. We anticipate that the silicate layer of Psyche can experience this compaction stage because of the applied pressure level. After this stage when the grains exceed the maximum compression index ($\sim$20 MPa), the number of crushed particles dramatically increases, leading to nearly 60$\%$ of the initial particles being broken and more than 60$\%$ of porosity being reduced. After porosity dramatically decreases, the particles finally approach the stabilizing stage - only 10$\%$ of the rest of particles are crushed even under the incremented loading pressure. These results are validated by comparing to previous experimental results performed by \cite{nakata2001one}. For Psyche, the pressures acting on the silicate-rich layer only range from 0 - 15 MPa, confirmed from our FEM simulation. Thus, within a few sub-kilometer, compaction is highly likely to begin by the rearrangement and breakage of a few particles as seen in the first stage in \cite{shi2016modeling}, although not significant compaction in very high pressure regime ($\sim 100$ MPa) is expected.  


Taking into account the compaction process on Psyche, we separate the silicate-rich layer into two parts: \textit{a top surface layer} which does not undergo the compaction process and \textit{a compressed surface layer} which is subject to the compaction. We set the threshold of the crushing limit as 10 MPa to differentiate those two layers. Of course, this threshold (10 MPa) is likely to be affected by grain features (i.e., grain shape, grain size distribution, and initial porosity of the structure). The initial state of grains may be critical to determine the crushing limit. From earlier studies on high-pressure compression tests of silica (e.g., \cite{hagerty1993one, de1958compressibility, nakata2001microscopic}), we notice some correlations between grain properties (e.g., initial relative density, particle angularity, and grain size distribution) and crushing behavior. However, we have hitherto insufficient information on the grain properties on the Psyche's surface, and thus it is impossible to figure out the accurate threshold of crushing limit for Psyche. Thus, we adopted the crushing limit of 10 MPa. To understand how our main result's trend depends on the threshold of the crushing limit, we further perform simulations by applying the lower and upper values of the crushing limit within the pressure region (0 - 15 MPa) in the silicate layer that is 5 MPa and 15 MPa, respectively.


\subsection{Shape model}
\label{subsec:shapemodel}
Currently, five 3D shape models of Psyche are available \citep{shepard2017radar,hanuvs2013sizes,kaasalainen2002models,viikinkoski201816, shepard2021asteroid}. Among them, we use the latest one by \cite{shepard2021asteroid}, which well constrained the topographic features of Psyche. This shape model is generated based on radar images obtained in 2005, 2015, and 2017 from the Arecibo S-band radar and in 2019 from the Atacama Large Millimeter Array, adaptive optics (AO) images in 2018 and 2019 from Keck and in 2015 and 2020 from the Very Large Telescope, and stellar occultations in 2010 and 2019. The dimension is estimated as 278 ($-4/+8$) $\times$ 238 ($-4/+6$)$\times$ 171 ($-1/+5$) in kilometers. Applying the nominal mass of 22.87 $\pm$ 0.70 $\times$ $10^{18}$ kg driven by \cite{baer2017simultaneous} and \cite{elkins2020observations} through the analysis of perturbations from other asteroids, \cite{shepard2021asteroid} estimated the overall bulk density of Psyche is $4.0 \pm 0.2$ g cm$^{-3}$.

\cite{shepard2021asteroid} assessed the major topographic features (i.e., dynamical depressions, craters, and missing mass region) found in their shape model, as well as those reported by other studies \citep{shepard2017radar,hanuvs2013sizes,kaasalainen2002models,viikinkoski201816}, in terms of whether they truly exist in Psyche or are artifacts of observation data processing. They reported two crater-like regions at the northern and southern hemispheres each and two missing mass regions at the equatorial region with convincing evidence, although other topographic features are still uncertain to be concluded. Among them, we note the crater-like regions named Panthia and Eros \citep{shepard2021asteroid}. Panthia is centered around longitude 300$^{\circ}$ and latitude -40$^{\circ}$ with a diameter of $\sim$90 km. \cite{shepard2021asteroid} suggested that this region possibly has a high metal concentration because its radar albedo ($>$ 0.4) is significantly higher than the background. This region is also detected by \cite{viikinkoski201816} and found to be much brighter (optically) than the surrounded areas. If Panthia is an impact crater, we anticipate that impact cratering processes \citep{hirabayashi2018role} likely mix the metallic materials in the core and the silicate layer to be eventually exposed on the surface at some levels. This scenario can partially explain why Panthia has the highest radar albedo, showing the localized metal concentration. Another certain crater, Eros, is located around longitude 290$^{\circ}$ and latitude -65$^{\circ}$ with a diameter between 50 and 75 km wide, which is smaller than Panthia. This area does not show high radar and optical albedos as Panthia has. However, the metal abundance in Eros cannot be ruled out because the area around Eros has bifurcated echoes suggesting multiple sources of high metal in the region \citep{shepard2021asteroid}. Eros is within one of those higher-albedo regions and might be the source of that stronger echo (M.K. Shepard, personal communication). In our analysis, we thus consider those two crater-like regions, Panthia and Eros, and estimate the core size that is possible to reveal the metallic core onto the surface via the impact crater process, assuming that the core shape is spherical.

\subsection{Dynamical environment}
Psyche's spin states are well constrained from earlier works  (e.g., \cite{kaasalainen2002models, vdurech2011combining, hanuvs2013sizes, shepard2017radar, shepard2021asteroid}) that showed all good agreements. We follow the estimates from \cite{shepard2021asteroid}. The spin period is $\sim$4.2 h with a spin pole at longitude $36^{\circ}$ and latitude $-8^{\circ}$. At present, there is no evidence indicating that Psyche is likely to be a non-principal axis rotator. Instead, given its large size and fast spin period, it is reasonable to consider that this body has a principal rotation behavior \citep{harris1994tumbling}.

\subsection{Study outline}
The primary goal of this study is to investigate the interior layout distribution assumed that Psyche is differentiated and constrain structural conditions consistent with the observed features. In Section \ref{sec:Methodology}, we describe how to develop an inverse problem algorithm to constrain the interior layout distribution of Psyche using the Three-layer model and FEM technique. The Three-layer model represents the structure of Psyche that consists of the iron core, the compressed layer, and the top surface layer. The FEM technique provides pressure distribution of the Three-layer model for structural analysis of Psyche. Section \ref{Sec:results} provides results obtained from the performed simulations, and in Section \ref{Sec:Discussions}, the results are discussed, mainly focusing on structural conditions that induce metal excavation via impact cratering process and ferrovolcanism. Finally, we argue the limitations that we noticed while using our current techniques in Section \ref{Sec:Limitations} and summarize our findings from this work in Section \ref{Sec:Conclusion}.
\section{Methodology} \label{sec:Methodology}
\subsection{Three-layer model}
\label{subsec:three-layer}
\begin{figure}[ht!]
	\centering
		\includegraphics[width=0.8\textwidth]{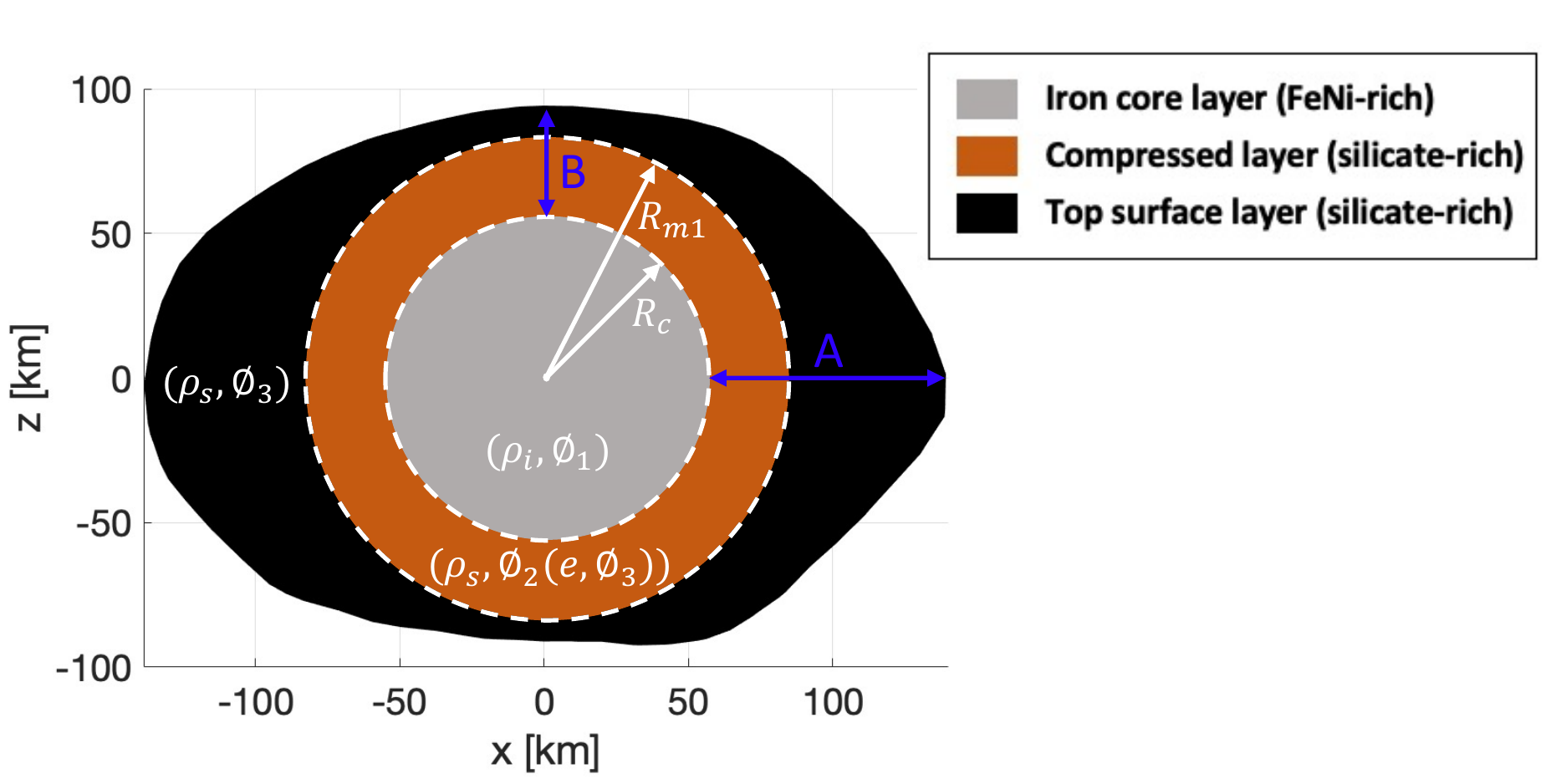}
	\caption{Three-layer model layout. Definitions of all symbols in the figure are described in Table \ref{tab:nomenclature}.}
	\label{FIG:THL}
\end{figure}

We develop a three-layer model that represents a structure of Psyche that contains an iron core, a compressed silicate-rich layer (compressed layer), and an uncompressed silicate-rich layer (top surface layer) (see Figure \ref{FIG:THL}). The entire shape is given by the radar-driven shape model \citep{shepard2021asteroid}. As noted in Figure \ref{FIG:THL}, the layouts of the iron core and the compressed layer are spherical as the compressed layer surrounds the core. The major motivation of this layer setting for the spherical layout assumption is that how the internal layers are distributed is highly unknown; potential issues of this assumption are discussed in Section \ref{Sec:Limitations} and thus subject to future investigation. The total bulk density of the structure is fixed at 4.0 g cm$^{-3}$ consistent with the earlier works that reported the Psyche's bulk density (e.g., \cite{elkins2020observations, viikinkoski201816, shepard2021asteroid}). However, each layer has the different bulk density as a combination of grain density ($\rho$) and porosity ($\phi$) depending on its composition. For example, the iron core has a denser density than the silicate-rich layer because of the abundance of metallic materials having higher grain density, and this structure is gravitationally stable. In the following section, we describe how to set the assumed ranges of grain density and porosity for each layer. The parameter setting are summarized in Table \ref{tab:properties}.

\begin{table}[ht!]
\renewcommand{\thetable}{\arabic{table}}
\centering
\caption{Nomenclature} \label{tab:nomenclature} \begin{tabular}{cl}
\hline
\hline
Symbol & Definition\\ \hline
$R_c$ & Core radius\\
$R_{m1}$ & Compressed layer radius\\
A & Minimum silicate layer thickness (lower bound)\\
B & Maximum silicate layer thickness (upper bound)\\
\hline
$\rho_{bulk}$ & Bulk density of Psyche\\
$V_{total}$ & Total volume of Psyche\\
$V_c$ & Volume of an iron core layer\\
$V_{m1}$ & Volume of a compressed silicate-rich layer\\
$\rho_i$ & Grain density of an iron meteorite (for the iron core layer)\\
$\rho_s$ & Grain density of a stony meteorite (for the silicate-rich layer)\\
$\phi_1$ & Porosity of an iron core\\
$\phi_2$ & Porosity of a compressed silicate-rich layer\\
$\phi_3$ & Porosity of a top surface layer\\
$e$ & Compression rate in the silicate layer (see Equation \eqref{eq14})\\
\hline
\end{tabular}
\tablecomments{All but $R_c$ and $R_{m1}$ are given parameters in our simulation.}
\end{table}

 \subsubsection{Grain density}
Grain density defines the bulk density of solid rocks that constitute the layer. In order to define this value, we refer to the bulk density of meteorites that have compositions similar to each layer. First, the silicate-rich layer is dominantly made up of silicate, consistent with the composition material of stony meteorites. Based on the meteorite data hitherto available, the bulk density of stony meteorites is well estimated between 2.5 and 3.5 g cm$^{-3}$ \citep{britt2003stony,carry2012density,consolmagno2008significance}. This range is also compatible with the bulk density of an S-type asteroid ($\sim$2.0 g cm$^{-3}$) \citep{fujiwara2006rubble} that mostly sustains macroporosity of around 40$\%$ \citep{abe2006mass} or less. Next, the metallic core layer is composed of iron (Fe) and Nickel (Ni). The best-fit meteorite whose mineralogy is dominated by iron and nickel is the iron meteorites. The grain density of this meteorite group depends on the Fe/Ni composition ratio (as the larger amount of Ni the meteorite has, the higher density is estimated), but be mostly distributed in the range from 7.0 to 8.0 g cm$^{-3}$ \citep{henderson1954discussion, britt2003stony}. 

\subsubsection{Porosity}
\label{subsec:porosity}
In this study, porosity defines macroporosity that measures the portion of pores in the layer. We assume that porosity is uniformly distributed in each layer for simplicity, although it may be a function of depth. This issue is discussed in Section \ref{Sec:Limitations}. Since the porosity of Psyche is poorly known, we consider it as a free parameter within a wide range to avoid any biased structural condition depending on this value. In order to determine the applicable porosity ranges, we refer to earlier studies (e.g., \cite{wieczorek2013crust, wilkison2000bulk,scheinberg2016core}) that analyzed porosity of well-explored celestrial bodies. 

For the iron core, we take into account two phases. The first phase is when the iron core was crystallized but never experienced fragmentation driven by significant impacts. If the core has not been subject to catastrophic impacts, it could sustain a relatively low porosity. \cite{scheinberg2016core} numerically investigated the inner core solidification process and showed porosity of the core could reach up to 5$\%$. However, if the body has been affected by huge impacts (the second phase), it could be fragmented, causing a relatively high porosity as seen in \cite{asphaug2006hit}. In general, large bodies such as Psyche may have porosity similar to lunar soils \citep{ostro1985mainbelt, elkins2020observations}. The porosity range of the lunar crust is estimated to vary from 4 to 21$\%$ to the depths of a few tens of kilometers, confirmed by feldspathic rocks obtained from Apollo missions \citep{wieczorek2013crust}. For now, it is impossible to precisely constrain the porosity range of the iron core of Psyche. Thus, we set an extensive range (0 - 30 $\%$) in simulations that can cover all ranges discussed above.

\begin{figure}[ht!]
	\centering
		\includegraphics[width=0.95\textwidth]{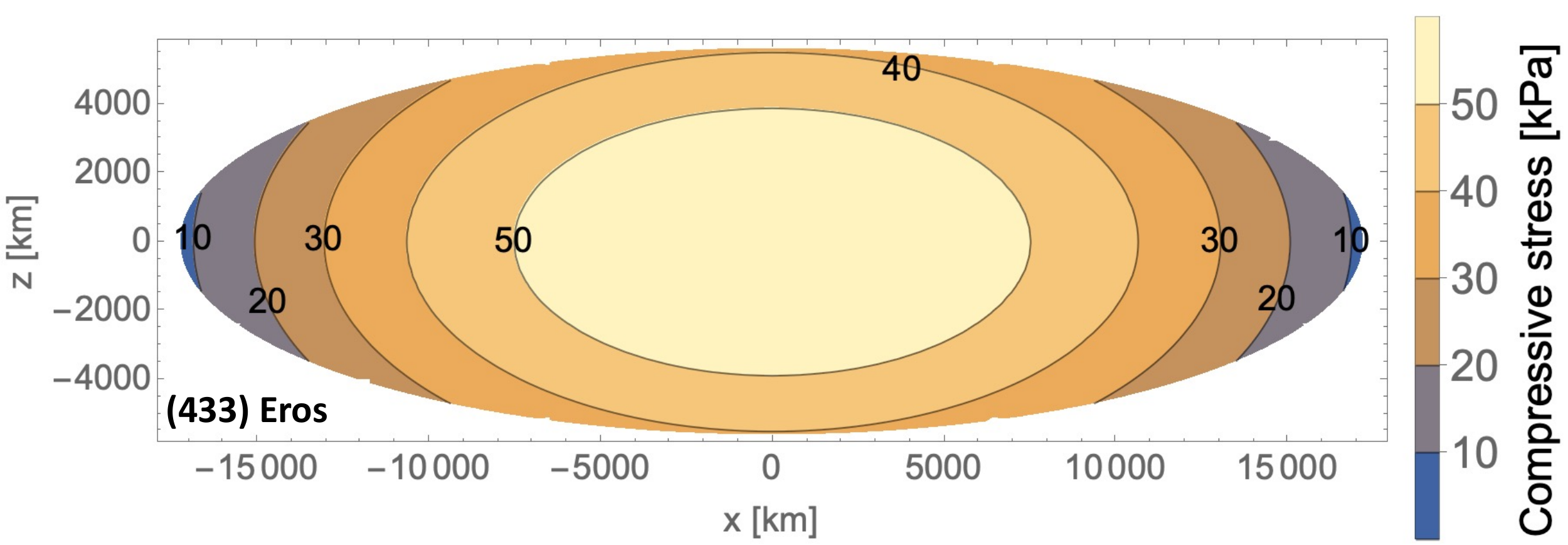}
	\caption{Elastic stress fields of Eros derived by a semi-analytical model \citep{nakano2020mass}}
	\label{FIG:SD_Eros_Itokawa}
\end{figure}

For the silicate-rich layer, we need to consider two layers - the top surface and the compressed layer. The top surface is defined as the layer where the pressure regime is less than the crushing limit (10 MPa) in our structure model. To infer the possible porosity range, we referred to the lunar crust and S-type asteroids given the similar bulk density and pressure regime with the top surface layer. Among large S-type asteroids, the well-examined object is Eros because the NEAR Shoemaker spacecraft has visited this body. The bulk density is given as 2.67 g cm$^{-3}$ by \cite{veverka2000near}, which is compatible with the determined grain density range (2.5 - 3.5  g cm$^{-3}$) of the silicate layer. This object is a shattered body, which means it may not be a rubble-pile but may simply be fractured, and has the bulk porosity ranging in 21$\%$ – 33$\%$ \citep{wilkison2000bulk, wilkison2002estimate}. We measured the central pressure of this object using a semi-analytical model, which provides elastic stress fields in a triaxial ellipsoid given its self-gravity and rotational force \citep{nakano2020mass}. Figure \ref{FIG:SD_Eros_Itokawa} shows the stress field of Eros when the shape dimension is assumed to be $34.4\times11.2\times11.2$ km, and the spin period is determined to be 5.27 h with a spin axis along the shortest principal axis (z-axis). As seen in the figure, the central pressure reaches tens of kilo-pascals ($>$ 50 kPa). This pressure level is located at the top surface of Psyche, under the depth of a few meters, possibly indicating that this location could sustain the similar bulk porosity of Eros. Although Eros implies that the very top surface of Psyche possibly has a high porosity, this does not provide any constraints on the deeper surface layer. We thus further look at the lunar crust that has been investigated using high-resolution gravity data measured from the dual Gravity Recovery and Interior Laboratory (GRAIL) spacecraft \citep{wieczorek2013crust,han2014global}. Earlier studies (e.g., \cite{wieczorek2013crust} and \cite{han2014global}) found that the lunar crust has the bulk density of 2.55 g cm$^{-3}$, consistent with the top surface layer, and the average porosity of 12 - 21 $\%$ to depths up to 15 km. We perform a simple calculation to compare the pressure regime of the lunar crust and Psyche. Given the zero-pressure at the lunar surface, a pressure (P) at a certain depth ($h$) can be computed as $\rho_{l}\times g_l \times h$. Here, $\rho_{l}$ and $g_{l}$ are the lunar crustal density of 2.55 g cm$^{-3}$ and the surface gravity of 1.6 m s$^{-3}$, respectively. Using this relation, we figure out that the depth of 2.4 km in the lunar surface has the pressure regime that matches the top surface layer of Psyche ($<$ 10 MPa). Given that pore closure likely occurs below 15 km depth in the lunar crust \citep{han2014global}, the top-surface layer of Psyche is still possible to sustain a similar porosity regime with the lunar crust. As a final note, it cannot be ruled out that the top surface layer of Psyche could be highly fragmented like a rubble-piles (e.g., Itokawa has bulk porosity over 40 $\%$ but a lower bulk density of 1.9 g cm$^{-3}$ than the top surface layer) or unconsolidated terrestrial sediments asteroid by impacts and re-accumulated process. Based on this, we set the very broad range of porosity of the top surface as 10 - 50 $\%$, which covers all porosity mentioned above.

Now, we describe the porosity range applicable in the compressed silicate-rich layer. The grain density of the compressed layer is the same as that of the top surface layer, but porosity decreases, as discussed in Section \ref{subsec:compaction}. \cite{shi2016modeling} numerically performed one-dimensional compression tests to study the particle breakage effect in silica sands and showed that the initial porosity decreases up to $\sim$30$\%$ at the compression pressure of 10 MPa. Given the compression rate, the porosity range of the compressed layer is set to be 7$\%$ - 35$\%$, which is the decreased range of the top silicate layer (10$\%$ - 50$\%$). However, one might question whether the 30$\%$ decrease is still significantly high to occur in the silicate layer under a low pressure regime of 10 MPa. To explore this issue, we additionally perform a simulation with a lower compression rate ($\sim5\%$) to understand how the compression rate influences the final structure layout. 

\begin{table}[ht!]
\renewcommand{\thetable}{\arabic{table}}
\centering
\caption{Geophysical parameter of each layer in the Three-layer model} \label{tab:properties} \begin{tabular}{ccc}
\hline
\hline
Layer& $\rho$ [$g cm^{-3}$] & $\phi$ [$\%$] \\ \hline
Top silicate-rich layer & 2.5 - 3.5 & 10 - 50 \\
Compressed silicate-rich layer & 2.5 - 3.5 & 7 - 35\\
Metal-rich layer & 7.0 - 8.0 & 0 - 30 \\
\hline
\end{tabular}
\tablecomments{$\rho$ represents a grain density.}

\end{table}

\subsection{Finite Element Model (FEM) Approach}
\label{Sec:FEM}
We develop a finite element model (FEM) to calculate the stress distribution of the layered structure of Psyche. This FEM is a modified version of the earlier model \citep{hirabayashi2021finite} to apply to the current problem for investigating a differentiated object. The major change made in the current version is considering density variations over a target body. For structural analysis, we start from the structural equation given as
\begin{linenomath}\begin{equation}\label{MEq1}
\rho\frac{\partial^2 \textit{u}}{\partial t^2} = \nabla^{T}\bm{\sigma} + \rho \textit{b} \approx 0
\end{equation}\end{linenomath}
where $\rho$ is the bulk density, $\textit{u}$ is the displacement vector that consists of [$u_x$, $u_y$, $u_z$], $\bm{\sigma}$ is the stress vector in the equilibrium state ($\rho\frac{\partial^2 \textit{u}}{\partial t^2} = 0$) that includes [$\sigma_{xx}$, $\sigma_{yy}$, $\sigma_{zz}$, $\sigma_{xy}$, $\sigma_{xz}$, $\sigma_{yz}$], and $\textit{b}$ is the loading vector that contains [$b_x$, $b_y$, $b_z$]. In the earlier model \citep{hirabayashi2021finite}, $\rho$ is considered as a constant parameter. However, in the current model, we apply the different value of $\rho$ depending on the layers. Equation \eqref{MEq1} eventually provides the stress field in the equilibrium state where the body is solely influenced by self-gravity and rotational forces. Thus, $\textit{b}$ is computed by a summation of gravity and centrifugal force. Then, the stress field is developed using Hook's principle as a constitutive law for linear elasticity. For the implementation, we expand this structural equation to a four-node tetrahedral FEM mesh that represents the structure of the target body. The detailed procedure is described in Appendix \ref{APX:FEM_general formulation}.

\subsubsection{FEM mesh}
We develop a 4-node FEM mesh using the ground-radar driven shape model by \cite{shepard2021asteroid}. The shape model is a surface mesh that consists of 1,652 vertices and 3,300 faces. To generate a tetrahedral volumetric mesh, we use TetGen, which is an open-source mesh generator developed by \cite{Hang2015}. The final mesh has 3,344 nodes and 15,569 elements. We accept this mesh quality because it does not induce any stress concentration caused when using a low-quality FEM mesh. In the FEM mesh, the elements are not isotropically distributed, but the interior tends to have relatively larger elements than those close to the surface layer. However, this element distribution is still reasonable to capture the stress fields distribution because the internal region has the least variation in the stress field \citep{hirabayashi2021finite}. In other words, the inside region of the structure is not much sensitive to the volume of elements.

\begin{table}[ht!]
\renewcommand{\thetable}{\arabic{table}}
\centering
\caption{Simulation input parameter settings} \label{tab:sims} \begin{tabular}{ccc}
\hline
\hline
Quantity & Value/Range & Units \\ \hline
Young's modulus & 1.0 $\times$ 10$^{7}$ & Pa \\
Poisson's ratio & 0.25 & - \\
Total bulk density & 4.0 & g cm$^{-3}$\\
Rotation period & 4.2 & h\\
Number of nodes & 3,344 & - \\
Number of elements & 15,569 & - \\
Shape dimension & 278 $\times$ 238 $\times$ 171 & km\\
\hline
\end{tabular}
\tablecomments{Young's modulus is an independent parameter of the stress solution in the linear-elastic deformation \citep{melosh1989impact}.}
\end{table}

\subsection{An inverse problem algorithm for structure layout}
\subsubsection{Structure layout constraints in the Three-layer model}\label{Sec:Methodology_size_constraints}
To estimate the size of the three layers (the core, the compressed, and the top surface layer), we use the geometric constraints with a constant total mass ($22.87 \times 10^{18}$ kg) of Psyche that induces a bulk density of 4.0 g cm$^{-3}$ given the current shape model \citep{baer2017simultaneous,elkins2020observations,shepard2021asteroid}. We then propagate the equation of the total mass of Psyche computed by the summation of each layer's mass as follows.

\begin{linenomath}\begin{equation}\label{eq13}
\rho_{bulk}V_{total} = V_{c}(1-\phi_1)\rho_{i} + V_{m1}(1-\phi_2)\rho_{s} + (V_{total}-V_c-V_{m1})(1-\phi_3)\rho_{s}
\end{equation}\end{linenomath}\label{PsycheCore}
where the variables are defined in Table \ref{tab:nomenclature}. Here, $\rho_{bulk}$, $V_{total}$ are observed quantities, $\rho_{i}$, $\rho_{s}$, $\phi_1$ and $\phi_3$ are assumed parameters. The details in the given parameter settings are described in Section \ref{subsec:porosity}. The rest of the parameters, including $\phi_2$, $V_c$, and $V_{m1}$, are driven by the following equations. 

\begin{linenomath}\begin{equation}\label{eq14}
\phi_2 = (1-e)\phi_3
\end{equation}\end{linenomath}

\begin{linenomath}\begin{equation}\label{eq15}
V_c = \frac{4}{3}\pi {R_c}^3
\end{equation}\end{linenomath}

\begin{linenomath}\begin{equation}\label{eq16}
V_{m1} = \int_{R_c}^{R_{m1}} V \,dV
\end{equation}\end{linenomath}
Here, $\phi_2$ is computed by $\phi_3$ and a compression rate ($e$) driven by the earlier compression tests of silica sands\citep{shi2016modeling,nakata2001one}. For example, if porosity at the compressed layer reduces 30$\%$, $\phi_2$ is defined as (1 - 0.3) $\times$ $\phi_3$ given e = 30 $\%$. Equation \eqref{eq15} and \eqref{eq16} show that $V_{c}$ and $V_{m1}$ are expressed by a function of the core radius ($R_c$) and the compressed layer radius ($R_{m1}$). Thus, Equation \eqref{eq13} is expressed as a function of $R_c$ and $R_{m1}$,  which are free parameters in the simulation.

\begin{figure}[ht!]
	\centering
		\includegraphics[width=0.5\textwidth]{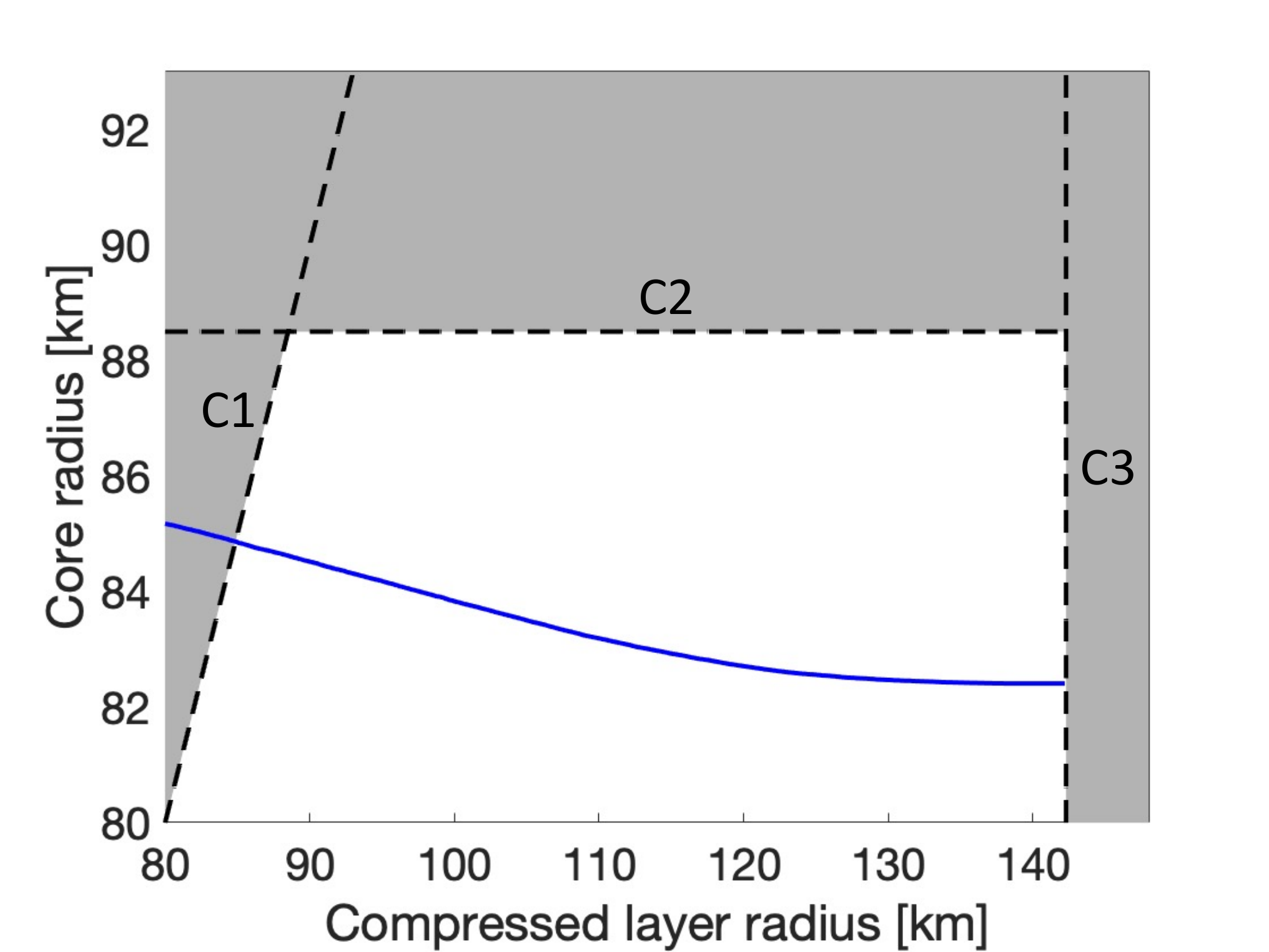}
	\caption{Correlation between a radius of the core ($R_c$) and the compressed layer ($R_{m1}$). The blue line represents a core radius depending on the compressed layer radius. The left dotted line (C1) is a boundary where the core radius is identical to the compressed layer. The upper horizontal line (C2) shows the minimum dimension ($\sim$88.5 km) along the shortest principal axis, while the right dotted line (C3) means the maximum dimension ($\sim$140 km) along the longest principal axis. The white area includes possible sets of ($R_c$, $R_{m1}$).}
	\label{FIG:SizeConstratins}
\end{figure}

Figure \ref{FIG:SizeConstratins} represents the correlation between $R_c$ and $R_{m1}$ (see the blue-colored line). Here, the white area is the considered region to pick a reasonable set of ($R_c$, $R_{m1}$), while the grey area is excluded. The boundary line (C1) is where the core radius is identical to the compressed silicate layer radius. The region below C1 indicates that the core is always placed below the compressed silicate layer. The upper horizontal line (C2) reflects the assumption of the spherical shape of the core. Thus, the maximum core radius is consistent with the minimum dimension (88.5 km) of Psyche along the shortest principal axis in the current shape model. Also, the line (C3) means the maximum dimension (140 km) of Psyche along the longest principal axis that the compressed layer radius cannot exceed. Based on this correlation, we can sort out possible sets of ($R_c$, $R_{m1}$) but still have many options. Thus, in order to further narrow down and determine one specific set of ($R_c$, $R_{m1}$), we incorporate our FEM technique.

\subsubsection{Structure layout constraints using FEM technique}
\begin{figure}[ht!]
	\centering
		\includegraphics[width=0.7\textwidth]{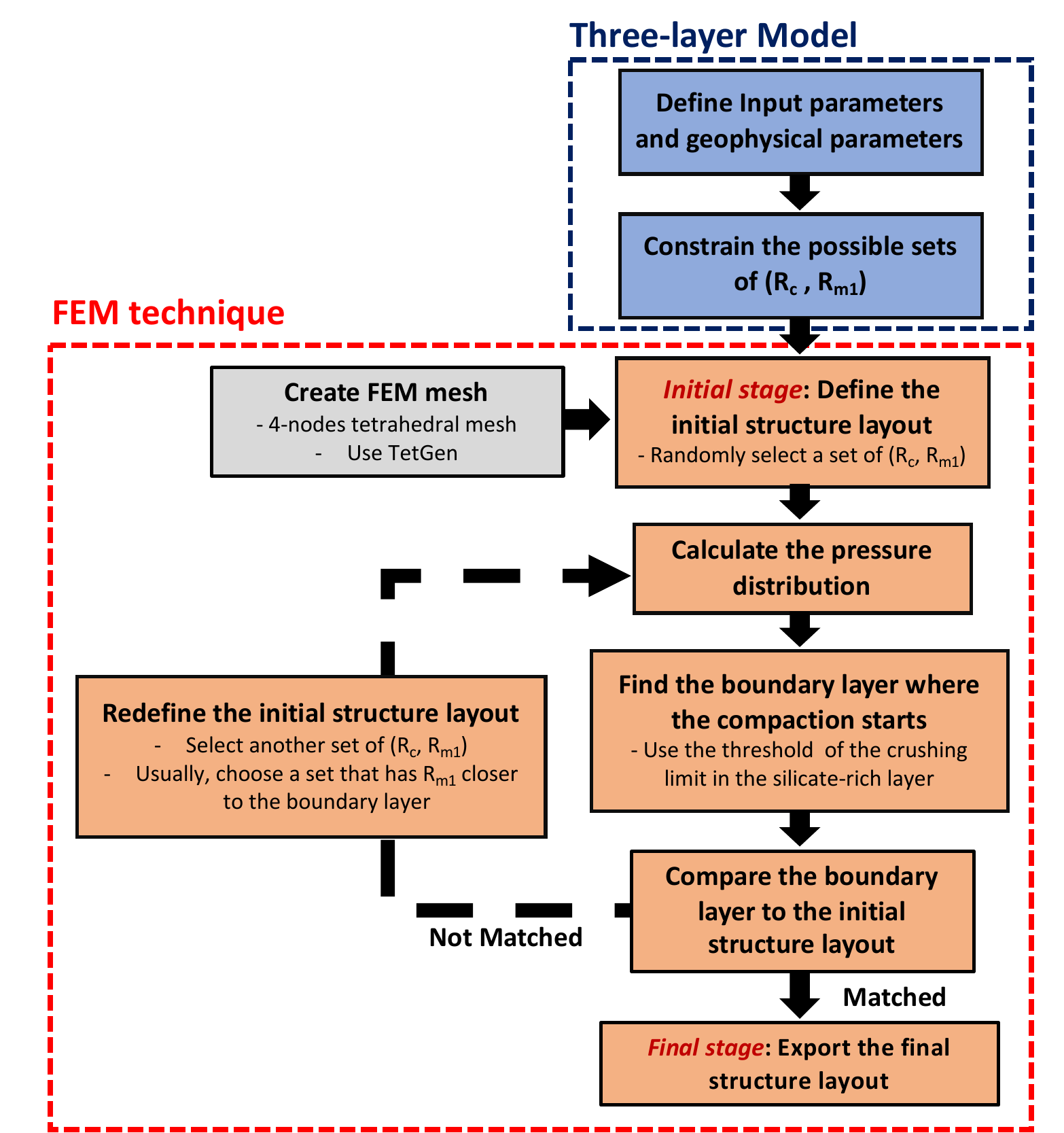}
	\caption{An inverse problem algorithm to constrain the structure layout using the Three-layer model and FEM technique. Currently, the bulk density and mass of Psyche have large uncertainties, although there is some supporting evidence inferring those parameters. Given this issue, we decided to fix the bulk density as 4.0 g cm$^{-3}$, which is the best-matched parameter from the earlier studies, in the Three-layer model. This value is estimated by \cite{shepard2021asteroid}, given the nominal mass is $22.87 \times 10^{18}$ kg which is driven by \cite{baer2017simultaneous} and \cite{elkins2020observations}. The matched case is defined before the final stage when the discrepancy between the boundary line and the compressed layer from the initial structure layout is less than 0.01 km.}
	\label{FIG:Algorithm}
\end{figure}

In order to find a unique structure layout at the given geophysical parameters, we adopt the condition where the compaction starts at the the threshold of the crushing limit in the silicate-rich layer. Using this condition, we can find out a unique case among the possible sets of ($R_c$, $R_{m1}$) from Equation \ref{eq13}. A numerical algorithm schematized in Figure \ref{FIG:Algorithm} describes the entire process of how to constrain the final structure layout of Psyche, and the FEM technique section shows the partial process of how to find the unique case among the possible sets of ($R_c$, $R_{m1}$) using the FEM technique developed in section \ref{Sec:FEM}.

\begin{figure}[ht!]
	\centering
		\includegraphics[width=1.0\textwidth]{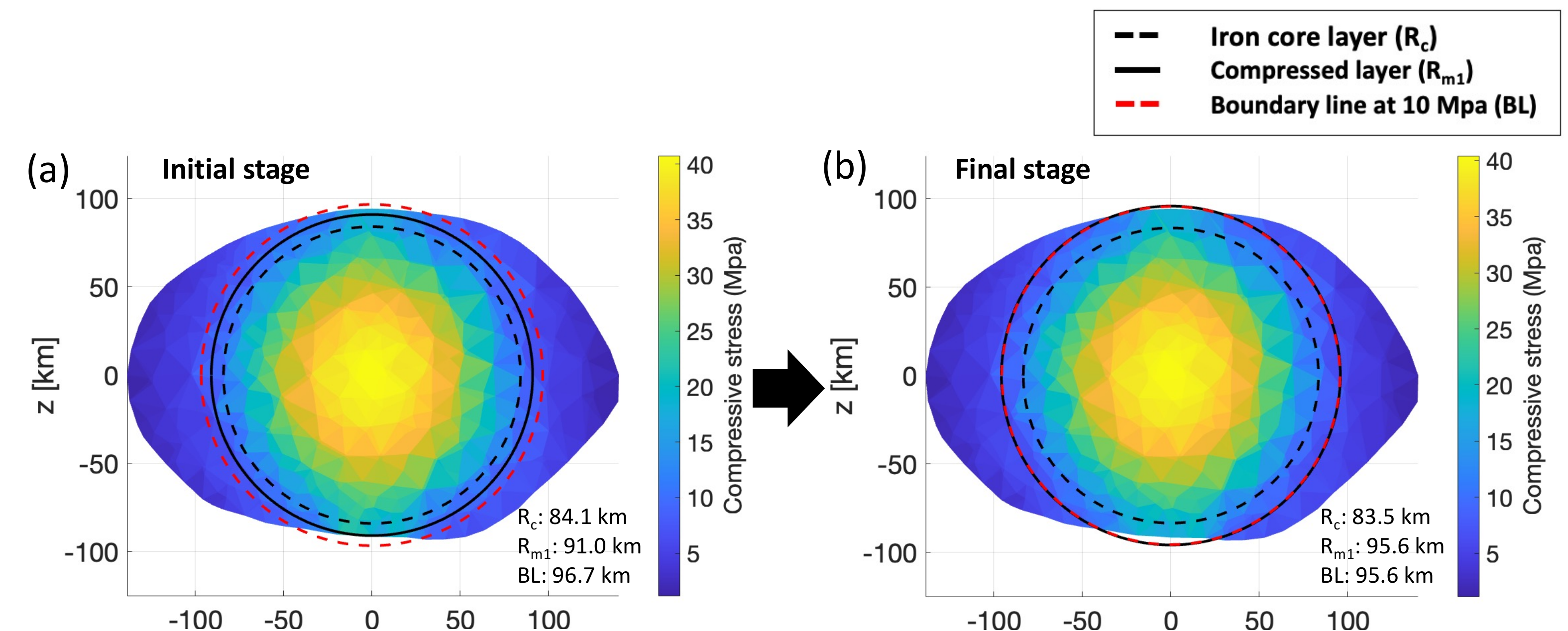}
	\caption{Structure layout (a) at the initial stage and (b) the final stage. Figure (b) show a unique case when the boundary line matches the crushing limit. Here, the crushing limit is set as 10 MPa. This case represents when the porosity of the iron-core and the top surface is 0$\%$ and 50$\%$ separately, given $e$ = 30 $\%$.}
	\label{FIG:MatchedCase}
\end{figure}

In the initial stage, we randomly select a set of ($R_c$, $R_{m1}$) among the possible ones and define it as the initial structure layout. Again, the possible sets are investigated based on the correlation between $R_c$ and $R_{m1}$ driven by Equation \eqref{eq13}, which defines the blue curve in Figure \ref{FIG:SizeConstratins}. We then compute the pressure distribution of the initial structure using FEM when this body rotates along the spin pole with a spin period of 4.2 h \citep{shepard2021asteroid}. After obtaining the pressure, we find the boundary line when the pressure reaches the threshold of the crushing limit in the silicate-rich layer. This boundary line indicates when the compaction process indeed starts in the given structure and must be consistent with the compressed layer. We check if the boundary line matches the compressed layer from the initial structure layout. If they do not match as seen in Figure \ref{FIG:MatchedCase} (a), we redefine the initial structure layout by applying another set of ($R_c$, $R_{m1}$). Usually, we use the set that has $R_{m1}$ closer to the boundary line because this condition makes the simulation converge faster. These processes, redefining the initial structure layout and computing the pressure distribution, are reiterated until the simulation converges into a unique case that gives the boundary line matched the compressed layer. In the final stage, we export the matched case as the final structure. Figure \ref{FIG:MatchedCase} (b) shows the final structure layout after the simulation ends. The input parameters are given in Table \ref{tab:sims} and the porosity of the iron-core and the top surface is set as 0$\%$ and 50$\%$ separately. We notice that both sets of ($R_c$, $R_{m1}$) used at the initial (Figure \ref{FIG:MatchedCase} (a)) and final stage (Figure \ref{FIG:MatchedCase} (b)) satisfy Equation \eqref{eq13} but the later one is the only case that provides the same pressure value as the crushing limit (this case, 10 MPa) at the compressed layer.

\section{Results}

\label{Sec:results}
We perform 7,500 simulations in total within the defined porosity ranges of the top surface layer and iron core with the crushing limit of 10 MPa (see Section \ref{subsec:result1_1}). Different porosity results in various bulk densities, leading to different core sizes and silicate-rich layer thicknesses. To find the dependence on the crushing limit of the silicate layer, we conduct additional simulations with the same simulation conditions but adopt a different crushing limit of silicates (see Section \ref{subsec:result1_2}). Lastly, we use a different compression rate (5 $\%$) in the compressed silicate layer and compare the result with the compression rate of 30 $\%$. Although we set the initial compression rate (30 $\%$) by referring to earlier studies (i.e., \cite{nakata2001one,shi2016modeling}) regarding one-dimensional compression test of silica, it may be different in the environment of Psyche depending on geological features such as grain size and shape.


\subsection{Constrained structure layouts of Psyche}
\label{subsec:result1_1}
\begin{figure}[h!]
	\centering
		\includegraphics[width=0.6\textwidth]{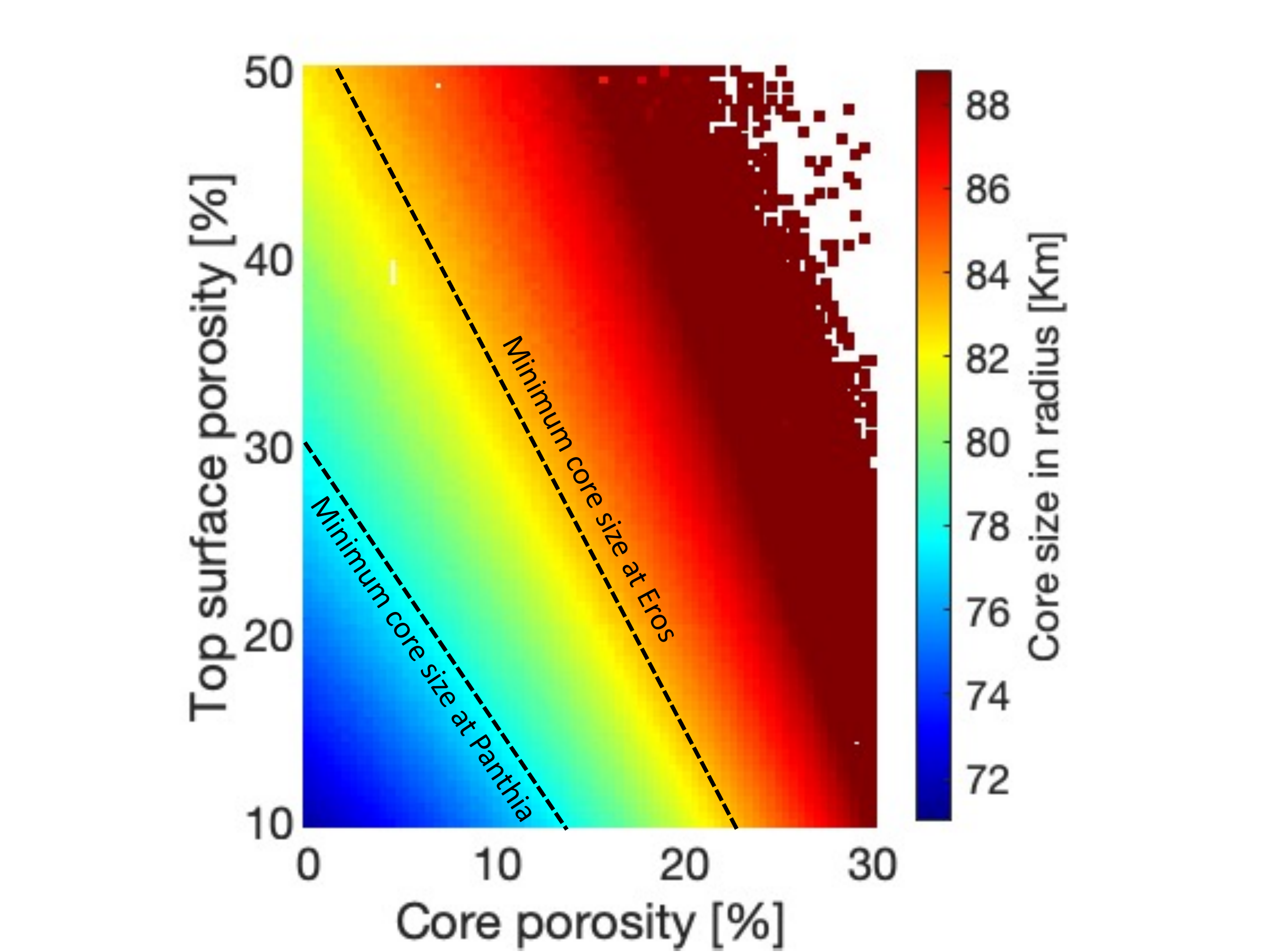}
	\caption{A colormap that shows the constrained core size within the possible porosity ranges. The black dotted lines define the minimum core size possible to be exposed at Panthia (Left) and Eros (Right) via an impact cratering process. In the determined porosity range, there are some cases at the bottom right corner where the porosity of the core is set to be higher than that of the overlying silicate layer. We address that this condition may not be realistic, although the composition discontinuity between the core and silicate layer may allow that porosity is not mutually related to each other.}
	\label{FIG:result1}
\end{figure}

\begin{figure}[h!]
	\centering
		\includegraphics[width=1.0\textwidth]{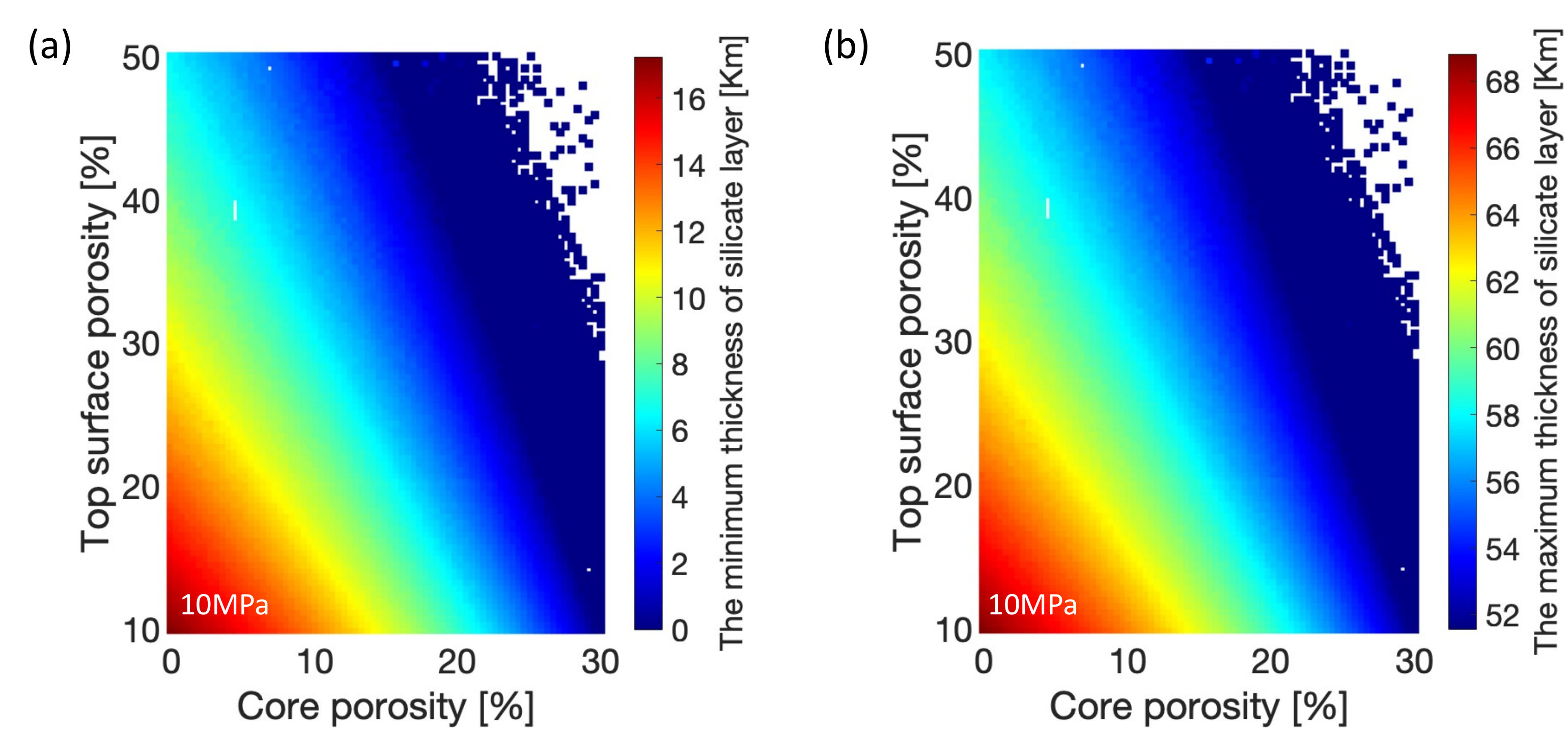}
	\caption{A colormap that shows the silicate-rich layer thickness: (a) The minimum thickness placed at the polar region and (b) the maximum thickness located at the equatorial region.}
	\label{FIG:10MPa_silicate}
\end{figure}

Figure \ref{FIG:result1} shows a color map representing the constrained core size in radius given the wide range of the core porosity and the top surface porosity. Here, We only consider the cases whose core sizes do not exceed the minimum dimension ($\sim$88.5 km) of the current shape model given the geometric constraint described in Section \ref{Sec:Methodology_size_constraints}. The trend seen in the color map entirely shows the core becoming larger as the core and silicate layer has a higher porosity. The correlation between volume and density accounts for this trend. As the more highly porous structure causes the lower bulk density, this body can sustain the larger volume under the same condition of mass. In Psyche, the large core eventually induces a relatively thin silicate layer. In detail, the estimate of core size varies from 72 km to 88.5 km. This core size takes up 30 - 45 $\%$ of the overall size of Psyche. The minimum core radius is located when the core and silicate layer have the lowest porosity, specifying the densest layers. Conversely, the largest core is placed at the most porous structure. Note that the current shape model shows an elongated shape having a minimum radius of 88.5 km along the shortest principal axis pointing at the pole and a maximum radius of 140 km along the longest axis heading the equator. Given this shape model, the thickest silicate-rich crust is place at the equatorial region, while the thinnest layer is located around the polar regions (represented as A and B in Figure \ref{FIG:THL}). Figure \ref{FIG:10MPa_silicate} shows the minimum and maximum thickness of the silicate layer that reaches $\sim$16 km and $\sim$68 km, respectively, when the structure is the densest. These minimum and maximum values can define the lower and upper limits of the silicate thickness. As a final note, we address that not all cases are converged. White pixels in the top-right corner seen in Figure \ref{FIG:result1} corresponded to the cases when a simulation did not converge. The non-convergence is the result of cases when the pressure in the silicate layer does not reach up to the crushing limit of 10 MPa, leading to no solution. The non-converged cases mostly occur when the core has a large porosity which is a very high upper limit for the porosity of the core, given that iron meteorites have porosity less than 10 $\%$ \citep{macke2010enstatite}. However, even if the overburden pressure in the silicate layer does not reach the crushing limit, this does not mean that Psyche cannot have a differentiated structure having a core and silicate layer. The structure layout is impossible to be determined in the inverse problem algorithm (Figure \ref{FIG:Algorithm}) because the boundary line does not exist but may simply be estimated by the bulk density of the silicate and core layer, which satisfy the bulk density of 4.0 g cm$^{-3}$ for the entire body, without considering the compaction effect.


\subsection{Dependence on the crushing limit of silicate}
\label{subsec:result1_2}
\begin{figure}
	\centering
		\includegraphics[width=1.0\textwidth]{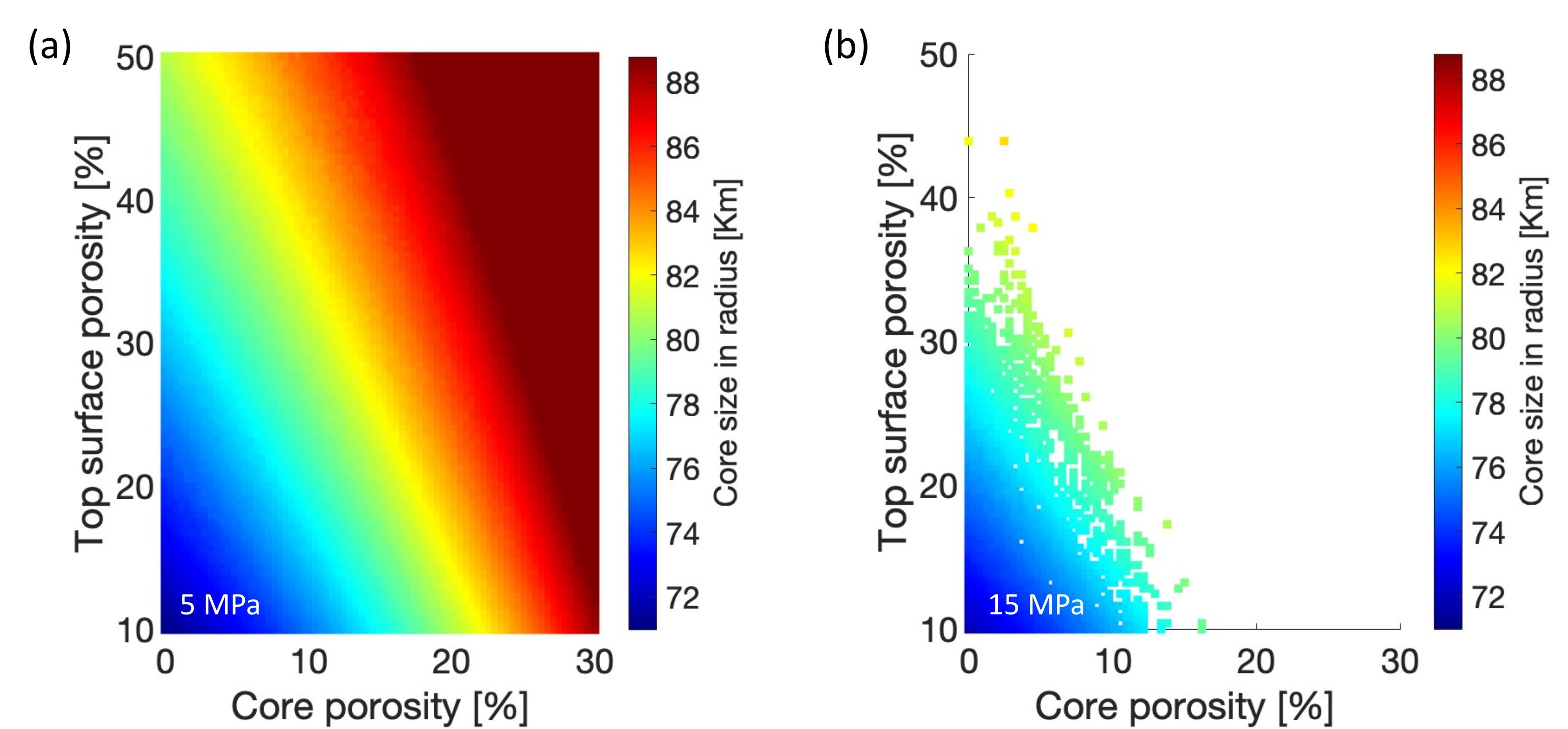}
	\caption{Comparison of colormaps between the crushing limit of (a) 5 MPa and (b) 15 MPa. In most cases where the core exceeds a certain size ($\sim$80 km), the 15 MPa provides a non-colored area that represents the simulation is not converged. We interpret this area as when the pressure in the silicate layer does not approach the defined crushing limit, and thus any solution for the interior layout does not exist.}
	\label{FIG:result2}
\end{figure}

\begin{figure}
	\centering
		\includegraphics[width=1.0\textwidth]{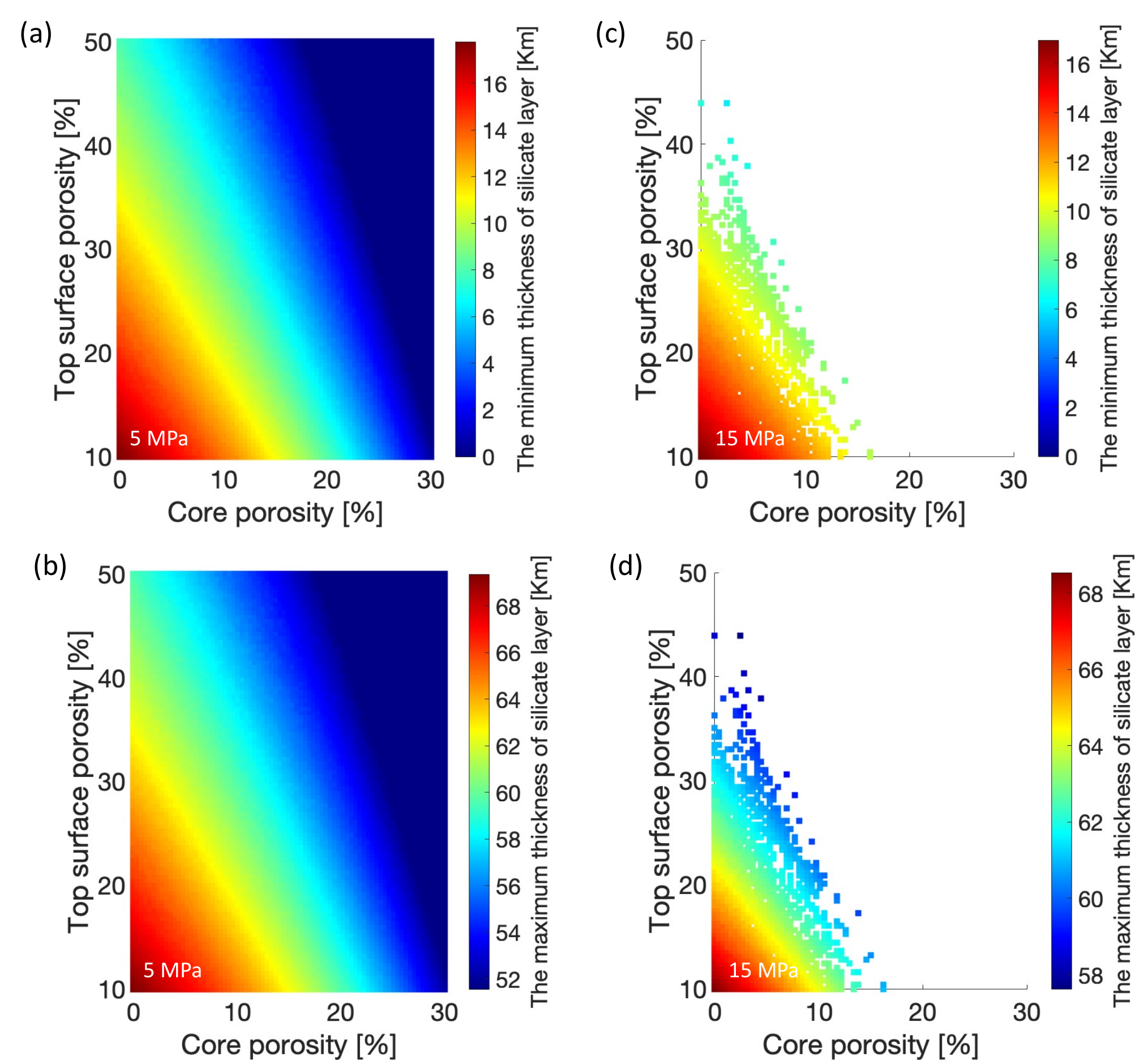}
	\caption{Comparison of colormaps that shows the silicate-rich layer thickness between the crushing limit of 5 MPa and 15 MPa: (a) The minimum thickness at 5 MPa, (b) the maximum thickness at 5 MPa, (c) The minimum thickness at 15 MPa, and (d) the maximum thickness at 15 MPa. As mentioned in Figure \ref{FIG:result2} (b), the 15 MPa provides a non-colored area that represents the simulation is not converged. Thus, the white area here also represents the case having no solution for the interior layout.}
	\label{FIG:comp_silicatelayer}
\end{figure}

We set the threshold of the crushing limit of silicate as 10 MPa by taking into account earlier studies (e.g., \cite{britt2003asteroid, shi2016modeling}). However, as mentioned in Section \ref{subsec:compaction}, this threshold can be controlled by grain properties (i.e., shape, size, initial porosity) on the surface of Psyche. Thus, we select two more different thresholds within the pressure regime ($\sim$15 MPa) of the silicate layer, the lower and upper values, and analyze how the trend of the core size and silicate layer thickness is affected. Figure \ref{FIG:result2} (a) and (b) show the core size in radius at the crushing limit of 5 MPa and 15 MPa, respectively. We first note that the 15 MPa case does not provide the solution in many simulation cases when the core size is larger than $\sim$80 km in radius, unlike the 5 MPa. This feature results from that the pressure regime in the silicate layer does not exceed 15 MPa, which is set as the crushing limit. Thus, the interior layout cannot be defined, while the 5 MPa and the 10 MPa cases find the solution under the same geophysical parameter settings. Except for the not-converging simulations, Figure \ref{FIG:result2} (a) and (b) capture the same trend seen in the previous section (Figure \ref{FIG:result1}). The core size becomes larger as the structure becomes porous. However, under the same condition of porosity, the core size converged becomes slightly different depending on the crushing limit, resulting in the shifted color distribution. For example, consider the simulation result at the point where the core and top-surface porosities are given as 0$\%$ and 30$\%$, respectively. The estimates of the core size are 76.6 km with the 5 MPa crushing limit, 77.9 km with 10 MPa, and 78.5 km with 15 MPa. With the lower crushing limit, the core size is estimated smaller. This is because compaction occurs closer to the surface in the silicate-rich layer with the lower crushing limit and thus induces the thicker compressed layer. This compressed layer has the denser bulk density due to a decrease in porosity and eventually takes up a large portion of Psyche's mass. Therefore, the smaller iron core exists as the compressed layer increases. Owing to this mechanism, given the lower crushing limit of 5 MPa, the smallest core size reaches 71.0 km in radius, while the 15 MPa case gives 71.8 km. We also note that the silicate layer thickness at 5 MPa slightly increases compared to the 10 MPa and 15 MPa cases (Figure \ref{FIG:comp_silicatelayer}). However, only a few hundred meters are changed in the core radius, and this difference can be negligible compared to the entire Psyche size.

\subsection{Dependence on the compression rate in the compressed layer} \label{subsec:result1_3}
\begin{figure}
	\centering
		\includegraphics[width=1.0\textwidth]{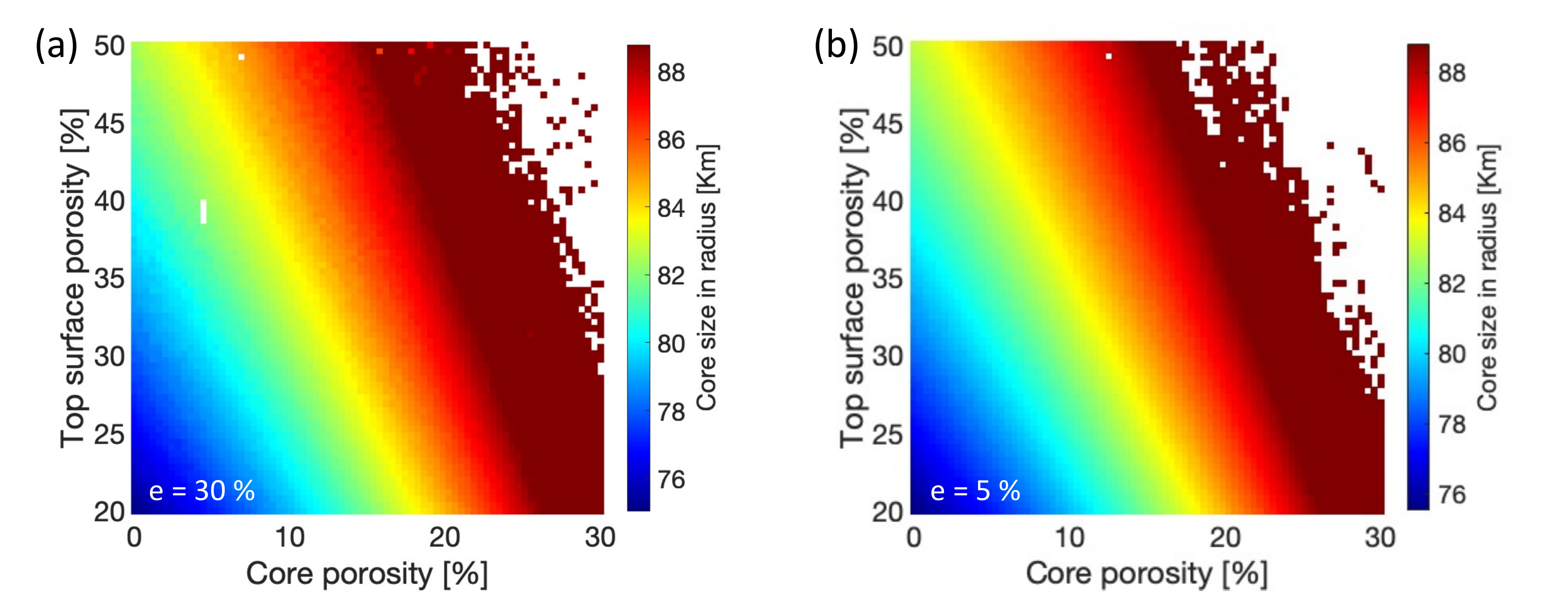}
	\caption{Comparison of colormaps that shows the core radius between the compression rate of 30 $\%$ and 5 $\%$}
	\label{FIG:comp_CompressionRate}
\end{figure}

One may question how the simulation results can change if the actual case of Psyche has a more minimal compression rate in the silicate layer. To understand the dependence on the compression rate, we perform supplementary simulations with the same simulation settings in Section \ref{subsec:result1_1} but a much smaller depression rate ($\sim 5\%$). Figure \ref{FIG:comp_CompressionRate} (a) and (b) show the core size at the compression rate of 30$\%$ and 5$\%$, respectively. This result shows that there is no noticeable difference between them. This is because even the highest compression rate of 30$\%$ is still small to reduce the bulk density of the compressed layer until affecting the core size. Also, since the compressed layer thickness is less thick than the top surface and core as seen in Figure \ref{FIG:MatchedCase} (b), the change in the compressed layer's bulk density does not have a significant effect on constraining the entire layout distribution. As a final note, we only run simulations within a reduced porosity range (20 - 50$\%$) for the top surface compared to the original setting (10 - 50$\%$), but this does not change our conclusion.

\section{Discussion}
\label{Sec:Discussions}
Based on the constrained interior layout distributions, we further analyze possible internal conditions that are compatible with the observed surface features of Psyche. Using the core size and the thickness of the silicate-rich layer defined by the simulations (Section \ref{subsec:result1_1}), we discuss the following likelihoods to reveal the metallic components in the core onto the surface if the differentiated structure is the case for Psyche: 1) If the iron core could be exposed at crater-like regions by the impact cratering process and 2) If Psyche could experience ferrovolcanic surface eruptions.

\subsection{Exposed metallic materials at crater-like regions}
\label{Sec:Discussion_1}
The current shape of Psyche \citep{shepard2021asteroid} exhibits topographic features that represent \textit{almost certain} craters at two locations as described in Section \ref{subsec:shapemodel}. Here, the confidence level - \textit{almost certain} - is determined in \cite{shepard2021asteroid} by comparing the current shape model to previous work \citep{shepard2017radar,viikinkoski201816,ferrais2020asteroid} in terms of topographical and optical albedo features. As described in Section \ref{subsec:shapemodel}, the radar albedo indicates potential metal abundances in those two crater-like regions. In the following discussion, we further constrain the compatible core size with the observed features on Panthia and Eros.

\begin{figure}
	\centering
		\includegraphics[width=1.0\textwidth]{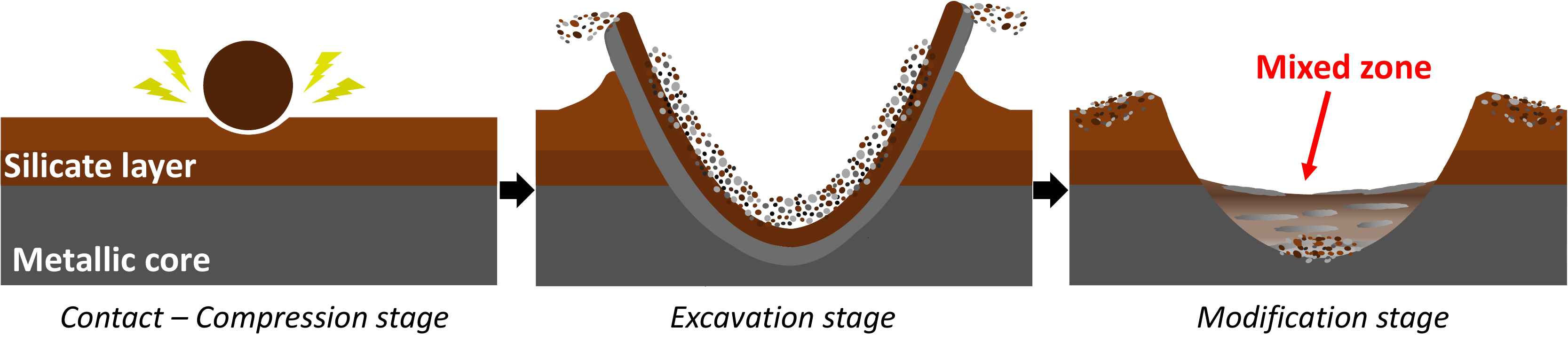}
	\caption{Impact cratering process schematic for a crater-like region on the surface of Psyche. Here, brown-like colors indicate silicates, while grey-like ones mean metal.}
	\label{FIG:Impactcrater}
\end{figure}

The process that likely causes the localized concentration of metals within Psyche's crater floor is impact cratering \citep{melosh1989impact,french1998traces,hirabayashi2018role}. In general, the impact crater formation has three stages: contact-compression, excavation, and modification stages. Figure \ref{FIG:Impactcrater} is a schematic depicting a plausible impact cratering scenario within the differentiated layers. After the contact-compression stage (when an impactor hits the surface and then the generated shock waves compress this area downward), a transient crater begins to form by ejecting away the materials (the excavation stage). If a transient crater formation penetrates a thin silicate layer to reach the metallic core, the ejected materials can be made up of mixed silicate and metal. The developed impact ejecta then radially falls onto the surrounding crater. The metal-mixed regolith via the impact-cratering procedure could explain the association of the crater-like region and its high radar albedo features observed in Psyche \citep{shepard2021asteroid}.

In order to constrain interior conditions to reveal the metals on the surface during an impact, we first estimate the maximum excavation depth ($H_{exc}$) where materials can excavate and become the ejecta. Based on the simple crater formation process, $H_{exc}$ is experimentally derived as only about one-third of the full depth of the transient crater, which is also approximated as one-tenth of the transient crater size ($D_t$) \citep{melosh1989impact}. The material deeper than $H_{exc}$ takes place beneath the transient crater floor, which cannot be revealed on the surface. Given the final crater sizes ($D_f$) of Panthia ($\sim$ 90 km) and Eros ($\sim$ 62.5 km), each of $D_t$ can be determined using a typical value ($\sim$ 1.2) of ${D_f}/{D_t}$ empirically derived for terrestrial craters \citep{melosh1989impact}. Eventually, the maximum excavation depth estimates are 10.8 km and 7.5 km for Panthia and Eros, respectively. The depths are a little deeper than their final crater depths, $\sim$ 10 km for Panthia and $\sim$ 4 km for Eros depicted in the current shape model. We then find the minimum core size that can be within the excavation areas by measuring the distance between the deepest excavated horizon and the center of the mass of Psyche. In Panthia, the core over the size of 78 km can reach the excavated zone, while Eros requires the larger one over 83 km. Panthia has the lower estimate of the core size to be exposed because its size is larger than Eros and thus induces a deeper excavated horizon. We marked this result as black dotted lines in Figure \ref{FIG:result1}. If a core size is over this line, we can infer that the metal in the core is ejected with the upper silicate layer in the excavation stage and eventually exposed at the breccia lens and ejecta blanket of the craters.

\subsection{Ferrovolcanic surface eruptions}
\label{Sec:Discussion_2}
Ferrovolcanism denominates the process that the core materials intrude into the covered rocky layer or even erupt onto the surface of planetesimals while solidifying. \cite{abrahams2019ferrovolcanism} first suggested that Ferrovolcanism may be an applicable mechanism to explain metallic components distributed on the surface of metallic asteroids. \cite{johnson2020ferrovolcanism} then analyzed that the differentiated Psyche may experience the ferrovolcanic surface eruption given the size and bulk density driven by \cite{drummond2018triaxial}. If this mechanism indeed occurs, there are a few requirements. First, Psyche's core should be crystallized from the surface to inside (i.e., inward solidification). If the core is too large, then the outward solidification is a more suitable process like the case for the Earth \citep{jacobs1953earth}. However, the estimated core size of Psyche, $\sim$100 km at maximum, is compatible with the inward solidification process \citep{haack1992asteroid,scheinberg2016core}. Second, the core should have melted sulfur-rich contents, which are lighter constituent elements than iron and nickel. Third, the core should produce a layer composed of solid iron-nickel and sulfur-rich liquid while solidifying  because the excess pressure generated by the density contrast between the solid and the melt becomes an energy source for melt to intrude the rocky layer. To address the second and third conditions on Psyche, \cite{johnson2020ferrovolcanism} referred to the simulation results regarding the inward core solidification conducted by \cite{scheinberg2016core}. This work confirmed that Psyche-sized core ($\sim$100 km in radius) is possible to produce pockets of sulfur-rich liquid within the solid iron-nickel while the core is cooling down. Using two free parameters, the vertical extent of melt and the amount of sulfur inside the melt, \cite{johnson2020ferrovolcanism} calculated the rocky layer thickness that the sulfur-rich melt can penetrate (Ref. figure 3 in \cite{johnson2020ferrovolcanism}). 

We set the minimum and maximum silicate layer thicknesses found in Figure \ref{FIG:10MPa_silicate} as the lower and upper limit of the silicate layer to assess the ferrovolcanic surface eruptions on Psyche. The minimum and maximum silicate layer reaches up to $\sim$16 km and $\sim$ 68 km when the structure has the lowest porosity. In the following description, we use the results from \citep{johnson2020ferrovolcanism} to suggest feasible conditions to experience the ferrovolcanic surface eruption given the silicate thicknesses. In order to penetrate the rocky layer of $\sim$ 16 km thick, the core needs to sustain the vertical extent of melts higher than at least 5 km, which can be sufficiently generated in the inward solidification of the Psyche-sized core \citep{johnson2020ferrovolcanism}. The required amount of sulfur-rich contents is at least $\sim$13 wt$\%$ S and reaches up to $\sim$27 wt$\%$ S as the vertical extent of melts decreases from 20 km to 5 km thick. On the contrary, the rocky layer of $\sim$ 68 km thick necessitates a much larger vertical extent of melts and sulfur-rich contents than the thinner case. The corresponding core condition to the thickness is at least 20 km height of the melt, including more than $\sim$25 wt$\%$ S of sulfur-rich contents. As assessed by \cite{johnson2020ferrovolcanism}, we also confirm that the defined structural layout still supports that Psyche might experience ferrovolcanism to reveal the metallic materials in the core onto the surface when the requirements (i.e., the size of the vertical extent of melts and amount of sulfur-rich contents) are satisfied. As a final note, we address that our results of the interior layout were obtained using the data (i.e., shape, rotation period, and bulk density) in the current state. However, Ferrovolcanim is a possible mechanism when the core is solidifying, which means that the timeline is unlikely to be the current but the early stage of the planetesimal formation. Therefore, the analysis above is only reasonable when Psyche has not experienced significant changes in its shape and physical properties compared to its primordial ones. 
\section{Limitations}

\label{Sec:Limitations}
In this section, we introduce two limitations in our current technique. 
\begin{enumerate}
    \item Mechanical properties of the structure may not be uniform in each layer.
    \item The iron core shape may not be a sphere but other shapes (i.e., an ellipsoid or irregular shape).
\end{enumerate}

The first issue means that the mechanical properties (bulk density and porosity) in our structure model are homogeneously distributed in each layer. In reality, these properties may change continuously as seen in the lunar crust \citep{besserer2014grail,wieczorek2013crust,han2014global}. For example, porosity likely decreases proportionally as the applied pressure is high. Likewise, the bulk density tends to be denser at the deeper layer. Thus, these mechanical properties are considered functions of depth in general. However, the currently available Psyche data is not sufficient to consider the details. This homogeneity assumption can be reconsidered as an inhomogeneous structure given its gravity field data obtained from the upcoming Psyche mission.

The second issue describes that we keep the core shape as a sphere. However, given that the gravity field of Psyche and its moment of inertia are currently unknown, this assumption can be reasonable. A recent study, \cite{ferrais2020asteroid}, suggested that Psyche might form at equilibrium, and its shape could be deformed by post-impacts. They addressed that the current shape presents small deviations from a Jacobi ellipsoid with a spin period of $\sim$3 hr and thus inferred that this might be Psyche’s primordial conditions. However, this work does not provide any constraints on the possibility of the core being in hydrostatic equilibrium. After the primordial shape had been formed, it is highly possible to have been influenced by some disturbances such as impacts \citep{asphaug2006hit} that might affect the internal structure. Given that there is no data (i.e., gravity field) to constrain its impact scenarios and primordial conditions, it is unavailable to model such a core shape of Psyche correctly. As with the first issue, we believe that the upcoming NASA's Psyche mission will provide more information (i.e., whether it is differentiated or undifferentiated, high-resolution topographic features to constrain its impact history) to better constrain its internal structure, and then we will revisit this issue.

\section{Conclusion}

\label{Sec:Conclusion}
In this study, we used the Three-layer model and FEM approach to constrain the interior layout of Psyche, given that this object is a metallic core covered with a silicate-rich layer. Below are the main results of our simulations.

\begin{itemize}
    \item If Psyche indeed originated from a differentiated planetesimal, given the reported bulk density of 4.0 g cm$^{-3}$, the estimated core size is 72 km to 88.5 km in radius (30 - 45 $\%$ of the overall size of Psyche) within the assumed porosity ranges. 
    
    \item The crushing limit of silicate affects the core size, but this change is minimal. With the lower crushing limit, the core size becomes smaller under the same geophysical condition because the compressed layer is thicker than the higher crushing limit case. Since this compressed layer has a high bulk density and takes up a more significant portion of Psyche's mass, the sustainable core size becomes smaller. The minimum core size reaches 71 km at the 5 MPa crushing limit while slightly increasing to 71.8 km at the 15 MPa case.
    
    \item The compression rate in the silicate layer has a minimal effect of constraining the core size. Under the pressure regime ( - 15 MPa) in the silicate layer of Psyche, compaction would occur with a less significant scale (compression rate less than 30 $\%$). Within this scale, the change in the compression rate does not induce any noticeable difference in the final interior layout.
    
    \item The current observation analysis shows that Panthia has the highest radar albedo than the surrounding, indicating a localized region of high metal concentrations. In addition, Eros cannot be excluded from metal abundance because this region is still within the area having elevated radar albedo and bifurcated radar echoes implying high metal sources. The minimum core size compatible with the impact cratering process to induce metal excavation at Panthia (Eros) is 78 km (83 km) in radius, which takes up 34 $\%$ (40 $\%$) of the total size of Psyche. The non-porous iron core even excavates its metal onto the surface at Panthia if the top silicate layer is highly fragmented ($>$ 30$\%$).
    
    \item The constrained interior layout has the minimum and maximum silicate layer of $\sim$16 km and $\sim$ 68 km at the pole and equator under the spherical core shape assumption when the structure has the lowest porosity. This structure condition still supports the ferrovolcanic surface eruption as suggested by \cite{johnson2020ferrovolcanism}. However, this interior layout is derived using Psyche's current physical conditions, which may not provide any insight into the timeline of Ferrovolcanism. Thus, this conclusion is only reasonable when Psyche still keeps its primordial conditions.
\end{itemize}

Our study suggested a possible interior structure of Psyche given the hypothesis that the asteroid is originated from a differentiated planetesimal. We then investigated the possible structural conditions compatible with the impact cratering process and ferrovolcanic surface eruptions to explain metal abundances on Psyche's surface. With detailed observations by NASA's Psyche in 2026, the current work will be extended to understand Psyche's internal structure, which will provide insight into its history.

\section*{Acknowledgement}
Y.K. and M.H. acknowledge support from Auburn University's Intramural Grant Program and Zonta Amelia Earhart Fellowship. We sincerely appreciate two anonymous reviewers for their assistance in improving this manuscript.

\bibliographystyle{aasjournal}

\appendix
\section{FEM general formulation}
\label{APX:FEM_general formulation}
In this section, we define a regular, bold and bold italic letter as a scalar, matrix and vector, respectively. We start with the structural equation in the form of a partial differential equation, driven by the theoretical assumption that all stress components in the continuum obey Newton's law of motion \citep{hughes2012finite}. Then we adopt Galerkin approximation that provides a numerical solution to a partial differential equation by finite element method \citep{oden2012introduction,johnson2012numerical}. The structural equation is given by
\begin{linenomath}\begin{equation}\label{Eq1}
\rho\frac{\partial^2 \textbf{\textit{u}}}{\partial t^2} = \nabla^{T}\bm{\sigma} + \rho \textbf{\textit{b}} 
\end{equation}\end{linenomath}
where $\rho$ is the bulk density, $\textbf{\textit{u}}$ is the displacement vector that consists of [$u_x$, $u_y$, $u_z$], $\bm{\sigma}$ is the stress vector that includes [$\sigma_{xx}$, $\sigma_{yy}$, $\sigma_{zz}$, $\sigma_{xy}$, $\sigma_{xz}$, $\sigma_{yz}$], and $\textbf{\textit{b}}$ is the loading vector that contains [$b_x$, $b_y$, $b_z$]. 
In the equilibrium state, the acceleration term, the left-hand side of Equation \eqref{Eq1}, becomes a zero ($\rho\frac{\partial^2 \textbf{\textit{u}}}{\partial t^2} = 0$). We then define the weak form of this equation by considering the four-node tetrahedral mesh. The weak form of Equation \eqref{Eq1} in the equilibrium state is given by
\begin{linenomath}\begin{equation}\label{Eq2}
\iiint_{V} (\Pi\nabla^{T}\bm{\sigma} + \Pi\rho \textbf{\textit{b}}) \dif V = 0
\end{equation}\end{linenomath}
where $\Pi$ becomes any given variable. 

We use a four-node tetrahedral FEM mesh and thus formulate this equation for each finite element. In the following discussion, the superscript $j$ represents the $j^{th}$ element. For example, $\textbf{\textit{u}}^j$ describes a 3-dimensional displacement vector of the $j^{th}$ solid element. Using the shape function ($\textbf{N}^j$), we further define $\textbf{\textit{u}}^j$ as $\textbf{N}^j\Bar{\textbf{\textit{u}}}^j$. Here, $\Bar{\textbf{\textit{u}}}^j$ is a 12-dimensional vector that consists of a 3-dimensional vector of each node of four-node tetrahedral elements. The shape function plays a role in approximately characterizing variables in the off-node region. The details are well described in Section 2.2.3 in \cite{hirabayashi2021finite}, and we follow the same definition of $\textbf{N}^j$. In the same way, we define $\textbf{\textit{b}}$ and $\bm{\sigma}$ as $\textbf{N}^j\Bar{\textbf{\textit{b}}}^j$ and $\textbf{N}^j\Bar{\bm{\sigma}}^j$, respectively. In the linear elasticity, we introduce Hooke's law to describe $\bm{\sigma}^j$, which is given as
\begin{linenomath}\begin{equation}\label{Eq3}
\bm{\sigma}^j = \textbf{K}^j \bm{\epsilon}^j 
         = \frac{E^j}{(1+\nu^j)(1-2\nu^j)}\begin{bmatrix} 
         1-\nu^j & \nu^j & \nu^j & 0 & 0 & 0 \\ 
         \nu^j & 1-\nu^j & \nu^j & 0 & 0 & 0 \\ 
         \nu^j & \nu^j & 1-\nu^j & 0 & 0 & 0 \\ 
          0 & 0 & 0 & \frac{1-2\nu^j}{2} & 0 & 0 \\
          0 & 0 & 0 & 0 & \frac{1-2\nu^j}{2} & 0 \\
          0 & 0 & 0 & 0 & 0 & \frac{1-2\nu^j}{2} \\
         \end{bmatrix}
         \quad
         \begin{bmatrix} 
         \epsilon_{xx} \\
         \epsilon_{yy} \\
         \epsilon_{zz} \\
         2\gamma_{xy} \\
         2\gamma_{xz} \\
         2\gamma_{yz} \\
         \end{bmatrix}
\end{equation}\end{linenomath}
where $E^j$ is Young's modulus, $\nu^j$ is Poisson's ratio, and $\bm{\epsilon}^j$ is the strain vector that consists of [$\epsilon_{xx}$, $\epsilon_{yy}$, $\epsilon_{zz}$, $\epsilon_{xy}$, $\epsilon_{xz}$, $\epsilon_{yz}$]. The components of the strain vector are replaced with $\nabla \textbf{N}^j \Bar{\textbf{\textit{u}}}^j$ as shown in the following logic.
\begin{linenomath}\begin{equation}\label{Eq4}
\bm{\epsilon}^j = \begin{bmatrix} 
             \frac{\partial u^j}{\partial x} \\
             \frac{\partial v^j}{\partial y} \\
             \frac{\partial w^j}{\partial z} \\
             \frac{\partial v^j}{\partial x} + \frac{\partial u^j}{\partial y} \\
             \frac{\partial w^j}{\partial x} + \frac{\partial u^j}{\partial z} \\
             \frac{\partial w^j}{\partial y} + \frac{\partial v^j}{\partial z} \\
             \end{bmatrix}
          = \begin{bmatrix} 
            \frac{\partial}{\partial x} & 0 & 0 \\
            0 & \frac{\partial}{\partial y} & 0 \\
            0 & 0 & \frac{\partial}{\partial z} \\
            \frac{\partial}{\partial y} & \frac{\partial}{\partial x} & 0 \\
            \frac{\partial}{\partial z} & 0 & \frac{\partial}{\partial x} \\
            0 & \frac{\partial}{\partial z} & \frac{\partial}{\partial y} \\
            \end{bmatrix}
            \quad
            \begin{bmatrix} 
            u^j \\
            v^j \\
            w^j \\
            \end{bmatrix}
          = \nabla \textbf{\textit{u}}^j = \nabla \textbf{N}^j \Bar{\textbf{\textit{u}}}^j.
\end{equation}\end{linenomath}
Substituting $\bm{\epsilon}^j$ in Equation \eqref{Eq3} into Equation \eqref{Eq4}, we then derive the relation, $\bm{\sigma}^j$ = $\textbf{K}^j \nabla \textbf{N}^j \Bar{\textbf{\textit{u}}}^j$.

Now, we can replace all vectors in Equation \eqref{Eq2} with that for the $j^{th}$ element. Here, $\Pi$ is given as $\textbf{N}^{jT}$. Then, the structural equation for the $j^{th}$ element is derived as
\begin{linenomath}\begin{equation}\label{Eq5}
\iiint_{V^{j}} \textbf{N}^{jT}\nabla^{T}(\textbf{K}^{j}\nabla \textbf{N}^{j}) \Bar{\textbf{\textit{u}}}^j \dif V^{j} + \iiint_{V^{j}} \textbf{N}^{jT} \rho^j \textbf{N}^{j} \Bar{\textbf{\textit{b}}}^j \dif V^{j} = 0
\end{equation}\end{linenomath}
Using the partial integral, we redefine the left-hand term as described in the below.

\begin{linenomath}\begin{equation}\label{Eq6}
\iiint_{V^{j}} \textbf{N}^{jT}\nabla^{T}(\textbf{K}^{j}\nabla \textbf{N}^{j}) \Bar{\textbf{\textit{u}}}^{j} \dif V^{j} = -\iiint_{V^{j}} (\nabla \textbf{N}^{j})^{T}\textbf{K}^{j}(\nabla \textbf{N}^{j}) \Bar{\textbf{\textit{u}}}^j) \dif V^{j}
\end{equation}\end{linenomath}
Then we finally derive the structural equation at the $j^{th}$ element. 

\begin{linenomath}\begin{equation}\label{Eq7}
\iiint_{V^{j}} (\nabla \textbf{N}^{j})^{T}\textbf{K}^{j}(\nabla \textbf{N}^{j}) \Bar{\textbf{\textit{u}}}^j) \dif V^{j} = \iiint_{V^{j}} \textbf{N}^{jT} \rho^j \textbf{N}^{j} \Bar{\textbf{\textit{b}}}^j \dif V^{j} 
\end{equation}\end{linenomath}

\begin{linenomath}\begin{equation}\label{ex}
f_g = \sum_{j=1}^{n_e} -\frac{G}{}
\end{equation}\end{linenomath}

The next step is to incorporate all the structural equations for each element into the entire body. The summation of $j^{th}$ elements at j = 1,2,...,$n_e$, where $n_e$ is the total number of elements in the FEM mesh, is defined as 

\begin{linenomath}\begin{equation}\label{Eq8}
\iiint_{V} f \dif V = \sum_{j=1}^{n_e} {\iiint_{V^{j}} f \dif V^{j}}
\end{equation}\end{linenomath}
where $f$ is an arbitrary function. Now we finally obtain the structural equation applicable to the four-node tetrahedral FEM mesh as the below. 

\begin{linenomath}\begin{equation}\label{Eq9}
\sum_{j=1}^{n_e} {\iiint_{V^{j}} (\nabla \textbf{N}^{j})^{T}\textbf{K}^{j}(\nabla \textbf{N}^{j})) \dif V^{j}}\Bar{\textbf{\textit{u}}} = \sum_{j=1}^{n_e} {\iiint_{V^{j}} \textbf{N}^{jT} \rho^j \textbf{N}^{j} \dif V^{j}} \Bar{\textbf{\textit{b}}}
\end{equation}\end{linenomath}
For simplicity, we further define Equation \eqref{Eq9} as 

\begin{linenomath}\begin{equation}\label{Eq10}
\bm{\rchi}\Bar{\textbf{\textit{u}}} = \bm{\Psi}\Bar{\textbf{\textit{b}}}
\end{equation}\end{linenomath}
where $\bm{\rchi}$ and $\bm{\Psi}$ are sparse matrix driven by
\begin{linenomath}\begin{equation}\label{Eq11}
\bm{\rchi} = \sum_{j=1}^{n_e} {\iiint_{V^{j}} (\nabla \textbf{N}^{j})^{T}\textbf{K}^{j}(\nabla \textbf{N}^{j})) \dif V^{j}}
\end{equation}\end{linenomath}
\begin{linenomath}\begin{equation}\label{Eq12}
\bm{\Psi} = \sum_{j=1}^{n_e} {\iiint_{V^{j}} \rho^j \textbf{N}^{jT}\textbf{N}^{j} \dif V^{j}}
\end{equation}\end{linenomath}

\subsection{Boundary conditions}
To solve the linear system in Equation \eqref{Eq10} with respect to the displacement, we need to estimate the inverse of $\Phi$, which is possible to cause singularity issues. We thus apply a proper boundary condition to our problem and then use an iterative conjugate gradient algorithm for the least-squares method to mitigate any singularity issues. For the boundary condition, we set the displacement at the center of mass (COM) of the body as zeros in any case. This boundary condition restricts any translational motions of the body able to be caused by loading. However, even if we apply this boundary condition, $\Phi$ is still a singular matrix, and thus we further need a mitigation process to reduce errors. For this, we adopt an iterative conjugate gradient approach for the least-square problem \citep[e.g.][]{hestenes1952methods} that is an efficient approach to solve the linear system with a large sparse matrix. It provides a unique solution by estimating the case when a residual is converged. The usage of this technique takes advantage when the $\Phi$ is a large and sparse matrix applicable to our current problem. Adopting this approach, we confirm that the residual is converged less than $\sim10^{-17}$, which is negligible.



\end{document}